\documentclass[journal]{IEEEtran}
\usepackage{graphicx,color}
\usepackage{cite}
\usepackage{setspace} 
\usepackage{amsmath}
\usepackage{amsmath,amsthm,amssymb}
\usepackage{multirow}
\usepackage{hhline}
\usepackage{epsfig}
\usepackage{subfigure}
\usepackage{epstopdf}
\usepackage{verbatim}
\usepackage{algorithm}
\usepackage{algorithmicx}
\usepackage{algpseudocode}
\usepackage{etoolbox}
\usepackage{cases}
\usepackage{csquotes}
\usepackage{enumitem}
\usepackage{mathtools,stmaryrd}
\SetSymbolFont{stmry}{bold}{U}{stmry}{m}{n}
\usepackage{bbm}
\usepackage{lettrine}
\usepackage{nicematrix}

\newcommand{\suchthat}{\;\ifnum\currentgrouptype=16 \middle\fi|\;}

\makeatletter
\newcommand*{\indep}{%
  \mathbin{%
    \mathpalette{\@indep}{}%
  }%
}
\newcommand*{\nindep}{%
  \mathbin{
    \mathpalette{\@indep}{\not}
  }%
}
\newcommand*{\@indep}[2]{%
  \sbox0{$#1\perp\m@th$}
  \sbox2{$#1=$}
  \sbox4{$#1\vcenter{}$}
  \rlap{\copy0}
  \dimen@=\dimexpr\ht2-\ht4-.2pt\relax
  \kern\dimen@
  {#2}%
  \kern\dimen@
  \copy0 
} 
\makeatother

\makeatletter
\newcommand*{\algrule}[1][\algorithmicindent]{%
  \makebox[#1][l]{%
    \hspace*{.2em}
    \vrule height .75\baselineskip depth .25\baselineskip
  }
}

\newcount\ALG@printindent@tempcnta
\def\ALG@printindent{%
    \ifnum \theALG@nested>0
    \ifx\ALG@text\ALG@x@notext
    \else
    \unskip
    \ALG@printindent@tempcnta=1
    \loop
    \algrule[\csname ALG@ind@\the\ALG@printindent@tempcnta\endcsname]%
    \advance \ALG@printindent@tempcnta 1
    \ifnum \ALG@printindent@tempcnta<\numexpr\theALG@nested+1\relax
    \repeat
    \fi
    \fi
}
\patchcmd{\ALG@doentity}{\noindent\hskip\ALG@tlm}{\ALG@printindent}{}{\errmessage{failed to patch}}
\patchcmd{\ALG@doentity}{\item[]\nointerlineskip}{}{}{} 
\makeatother

\newtheorem{theorem}{Theorem}

\newtheorem{corr}{Corollary}

\allowdisplaybreaks

\begin{document}

\title{DCSK-based Waveform Design for Self-sustainable RIS-aided Noncoherent SWIPT}

\author{Authors}
\author{Priyadarshi Mukherjee, \textit{Senior Member, IEEE}, Constantinos Psomas, \textit{Senior Member, IEEE},\\ and Ioannis Krikidis, \textit{Fellow, IEEE}
\thanks{P. Mukherjee is with the Advanced Computing \& Microelectronics Unit,  Indian Statistical Institute, Kolkata, India. C. Psomas and I. Krikidis are with the Department of Electrical and Computer Engineering, University of Cyprus, Nicosia 1678 (E-mail: priyadarshi@ieee.org, psomas@ucy.ac.cy, krikidis@ucy.ac.cy).

This work has received funding from the European Union's Horizon-JU-SNS research and innovation programme under the project iSEE-6G (Grant agreement No. 101139291). It has also received funding from the European Research Council (ERC) under the European Union's Horizon 2020 research and innovation programme (Grant agreement No. 819819).
}}

\maketitle

\begin{abstract}
This paper investigates the problem of transmit waveform design in the context of a chaotic signal-based self-sustainable reconfigurable intelligent surface (RIS)-aided system for simultaneous wireless information and power transfer (SWIPT). Specifically, we propose a differential chaos shift keying (DCSK)-based RIS-aided point-to-point set-up, where the RIS is partitioned into two non-overlapping surfaces. The elements of the first sub-surface perform energy harvesting (EH), which in turn, provide the required power to the other sub-surface operating in the information transfer (IT) mode. In this framework, by considering a generalized frequency-selective Nakagami-$m$ fading scenario as well as the nonlinearities of the EH process, we derive closed-form analytical expressions for both the bit error rate (BER) at the receiver and the harvested power at the RIS. Our analysis demonstrates, that both these performance metrics depend on the parameters of the wireless channel, the transmit waveform design, and the number of reflecting elements at the RIS, which switch between the IT and EH modes, depending on the application requirements. Moreover, we show that, having more reflecting elements in the IT mode is not always beneficial and also, for a given acceptable BER, we derive a lower bound on the number of RIS elements that need to be operated in the EH mode. Furthermore, for a fixed RIS configuration, we investigate a trade-off between the achievable BER and the harvested power at the RIS and accordingly, we propose appropriate transmit waveform designs. Finally, our numerical results illustrate the importance of our intelligent DCSK-based waveform design on the considered framework.
\end{abstract}


\IEEEpeerreviewmaketitle

\section{Introduction}
In recent years, wireless traffic has been exponentially increasing due to applications such as massive machine-type communications and Internet of Things (IoT). Precisely, with the fifth Generation (5G) mobile subscriptions expecting to exceed $5.3$ billion in $2029$, the global IoT connections is forecast to reach around $38.9$ billion in $2029$ \cite{ericsson}. Therefore, with the 5G struggling to cope up with these challenges of scalability and rising sustainability issues, several research initiatives on the next generation technology, i.e., 6G,  have been launched around the world \cite{psomas6g}. This implies, that in scenarios, where large number of wireless devices are spread over a geographical region, the aspect of powering or charging them is a very critical issue. Therefore, the aspect of self-sustainability in wireless communication networks is gaining both relevance and importance in industry as well as academia. The utilization of renewable energy sources such as solar energy and the harvesting of ambient radio frequency (RF) energy represent viable options for the purpose of improving the aspect of self-sustainability of these networks. However, due to the uncontrollability and erratic nature of these alternatives, the possibility of harvesting energy from controlled RF sources, i.e., the concept of wireless power transfer (WPT) is appearing to be an attractive and viable solution \cite{psomas6g}. In this context, the possibility of simultaneous wireless information and power transfer (SWIPT) by the RF signals is also gaining attention and being considered as a promising technology \cite{psomas6g}.

On the other hand, reconfigurable intelligent surfaces (RISs) have been recently proposed to `control' the wireless propagation environment \cite{risov}. An RIS is essentially an array of reconfigurable passive elements embedded on a flat metasurface, which controls the propagation medium instead of adapting to it \cite{ris1}. This is the basic difference between RIS and other technologies (such as adaptive modulation and beamforming) developed over the last decade, where the unifying motivation is to intelligently adapt to the  wireless channel instead of having a control over it.

\subsection{Literature Review}
With the RIS simply reflecting the incident signal in a desired direction, it does not require any dedicated RF chains and the operation takes place simply by tuning the parameters of its passive elements \cite{ris2}. In terms of power consumption, a finite amount of energy is consumed by these passive elements for their reconfigurability, which is significantly lower as compared to traditional relays \cite{epower}. This factor is responsible for the RISs being termed as passive or `nearly' passive structures. Based on these considerations, the concept of self-sustainable RIS has recently been proposed \cite{zeris1,zeris2,zeris3,zeris}. The work in \cite{zeris1} proposes a self-sustainable RIS powered hybrid relaying scheme to enhance the performance of wireless powered communication networks. The authors in \cite{zeris2} investigate the advantages of an RIS-enabled WPT system and propose an efficient RIS-aided WPT scheme. The work in \cite{zeris3} discusses the feasibility of converting a conventional RIS to a self-sustainable entity by incorporating the aspect of energy harvesting from the incident signals. The authors in \cite{zeris} propose suitable characterization metrics for quantifying the performance of a self-sustainable RIS-aided system with multiple energy harvesting (EH) modes. Therefore, we observe that this particular class of RISs is designed to operate with minimal or no external power by harvesting energy from the incident signals, unlike a conventional RIS that typically relies on a dedicated power supply. Hence, in the context of 6G communication networks, it is ideal for scenarios with limited power supply or where autonomous operation is beneficial, offering a sustainable and efficient solution.

Note that, all the above-mentioned works consider a linear EH model at the RIS. While this is an over-simplified model of the EH circuit, there are other works that model the EH circuit more accurately. Specifically, based on the saturation of the output power beyond a certain RF input power due to diode breakdown, the studies in \cite{plinear} and \cite{satm} propose a piece-wise linear and a tractable logistic nonlinear model of the EH circuit, respectively. The logistic model, which is obtained by fitting measurements from a practical EH circuit, is basically an improved version of the above-mentioned linear and piece-wise linear models. The authors in \cite{ehrev} characterize the power conversion efficiency of the EH circuit as a a rational function of the average input power. In this direction, the work in \cite{comb} proposes a realistic EH circuit characteristics dependent nonlinear EH model, which also incorporates the aspect of transmit waveform design to maximize the EH conversion efficiency. The experimental studies in \cite{chaosexp1} demonstrate that signals with high peak-to-average-power-ratio (PAPR), like the chaotic waveforms, provide higher WPT efficiency. The authors in \cite{wcl2} investigate a continuous-time chaotic signal-based WPT framework, which corroborates the above observation.

Generally, chaotic signals are extensively used for the purpose of information security and wireless privacy \cite{sec2}. However, as practical operating platforms cannot support the continuous nature of the system variables and the solutions of differential functions cost hardware resources, the discrete-time chaotic system  appears to be an interesting alternative. As a result, the differential chaos shift keying (DCSK) is one of the widely investigated discrete-time chaotic
signal-based communication techniques \cite{text}. Moreover, DCSK is also noncoherent in nature and it is a transmitted reference modulation technique, comprising of a reference frame and an identical/inverted replica of the same depending on the data transmitted. For enhancing the system data rate, the work in  \cite{srdcsk} develops a shorter symbol duration-based DCSK system, namely, the short reference DCSK (SR-DCSK). In this direction, there are some works that aim to exploit the benefit of both WPT/SWIPT and DCSK \cite{jstsp,chaoswipt3,chaoswipt4,chaoswipt5,twc}. By considering the nonlinearities of the EH process, the authors in \cite{jstsp} investigate DCSK-based transmit waveform designs to improve the WPT efficiency. The authors in \cite{chaoswipt3} demonstrate the impact of index modulation and its transmission characteristics on a basic DCSK-based SWIPT set-up, which results in reduced energy consumption. The work in \cite{chaoswipt4} proposes an adaptive link selection scheme for buffer-aided relaying in a decode-and-forward relay-based DCSK-SWIPT framework. Particularly suitable for low-power and low-cost e-health IoT applications, the work in \cite{chaoswipt5} propose a multiple-input single-output SWIPT architecture based on code index modulated $M$-ary DCSK. In \cite{twc}, the authors analyse a DCSK-based SWIPT multiantenna receiver architecture, where each antenna, depending on the application requirement, switches between information transfer and EH modes. Moreover, there are a few very recent works, that explore the advantages of RIS in DCSK-based WPT/SWIPT \cite{dcskris1,dcskris2,dcskris3}. The authors in \cite{dcskris1} propose a linear EH model dependent RIS and energy buffer aided DCSK-based wireless powered communication system, where a simple time-switching scheme is designed to minimize the associated bit error rate (BER). By considering the logistic nonlinear EH model, the work in \cite{dcskris2} investigates an RIS-assisted frequency-modulated DCSK-based SWIPT architecture. Here, the authors specifically consider two deployment scenarios, namely, the RIS-assisted access point and dual-hop communication. In \cite{dcskris3}, a non-coherent RIS-aided $M$-ary differential
chaos shift keying system is proposed, which offers superior BER performance without the channel state information (CSI). However, the aspect of chaotic waveform-based transmit waveform design for self-sustainable RIS-aided SWIPT and its corresponding analytical characterization have not been explored.

Besides, the work in \cite{zeris} characterizes the self-sustainable RIS-aided systems in terms of `the joint energy-data rate outage probability', which is defined as the union of the energy outage event and the data rate outage event. Here, the conventional Shannon capacity is used as the measure of data rate, i.e., the underlying assumption is that of having very lengthy codewords with the associated BER asymptotically going to zero. On the other hand, chaotic signal-based codewords are finite length in nature, which implies that the above characterization, in this case, is not valid. Hence, we investigate the BER-harvested power trade-off for such systems and its impact on the configuration of the RIS and the associated transmit waveform design.

\begin{table*} [!t] 
\begin{center}
  \caption{Summary of notations.}
  \vspace{-2mm}
\resizebox{0.92\textwidth}{!}{%
{\renewcommand{\arraystretch}{2} 
  \begin{tabular}{|c|c||c|c|}
    \hline
    \bf{Notation} & \bf{Description} & \bf{Notation} & \bf{Description} \\
    \hline\hline
    $\beta$ & Spreading factor &
    $N$ & Total number of reflecting elements in the RIS \\
    \hline
    $d_l$& $l$-th information bit &$M(K)$ & Number of reflecting elements working in the IT (EH) mode\\
    \hline
     $\phi$& Reference length& $x_{l,k}$ & $k$-th component, $l$-th transmission interval chaotic sequence\\
    \hline
    $\varphi_m$ & Reflection coefficient of  $m$-th RIS element & $s_{l,k}$ & $k$-th component, $l$-th transmission interval transmitter output \\
    \hline
    $P_{\rm inf}$ & Power consumption of each reflecting RIS element & $L_{sr}(L_{rd})$& Number of independent paths in Tx-RIS (RIS-Rx) channel \\
    \hline
    $P_{\rm cont}$ & RIS controller
power consumption &$h_{p,n}$ & $p$-th path, $n$-th RIS element complex Tx-RIS channel coefficient\\
    \hline
    $C_0$ & Pathloss at one meter distance & $\alpha_{p,n},\theta_{h,p,n}$ & Amplitude and phase of $h_{p,n}$, respectively, with $\mathbb{E}\{\alpha_{p,n}^2\}=\Omega_{\alpha,p,n}$ \\
    \hline
    $P_{\rm t}$ & Transmission power & $\alpha_{sr}(\alpha_{rd})$ & Path-loss exponent of Tx-RIS (RIS-Rx) channel \\
    \hline
    $d_{sr}$ $(d_{rd})$& Tx-RIS (RIS-Rx) distance & $g_{n,q}$ & $n$-th RIS element, $q$-th path complex RIS-Rx channel coefficient\\
    \hline
    $w_{sr,n}(w_{r,d})$ & AWGN at $n$-th RIS patch (Rx) & $\beta_{n,q},\theta_{g,n,q}$ & Amplitude and phase of $g_{n,q}$, respectively, with $\mathbb{E}\{\beta_{n,q}^2\}=\Omega_{\beta,n,q}$ \\
    \hline
    $\Phi$ & Phase shift matrix & $y^{\rm R}_{l,k,n}$ & Received signal at $n$-th RIS element, for transmit signal $s_{l,k}$\\
    \hline
    $\theta_{e,m}$ & Phase error at $m$-th element of RIS IT section & $\boldsymbol{y_{l,k}^{\rm R}}$ & Received signal vector at the RIS, for transmit signal $s_{l,k}$ \\
    \hline
    $\lambda_l$ & Decision variable for $l$-th information bit & $\boldsymbol{y_{l,k}^{\rm R,I}}(\boldsymbol{y_{l,k}^{\rm R,P}})$ & Received signal vector at the RIS IT (EH) section \\
    \hline
    $y_{l,k}^{\rm Rx}$ & Received signal at Rx, for transmit signal $s_{l,k}$  & $y^{\rm R}_{l,k,i}$ & Received signal at $i$-th patch of RIS, for transmit signal $s_{l,k}$ \\
    \hline
  $\nu_1,\nu_2,R_L$ & Harvesting circuit parameters & $y_{l,i}^{\rm AC}$ & Analog correlator output at $i$-th element of RIS EH section,  for information bit $d_l$ \\
    \hline
    $v_{i,{\rm out}}$ & Output DC voltage at $i$-th element of RIS EH section & $P_{\rm harv} $& Total harvested DC power in RIS EH section \\
    \hline
    $E_{\rm b},T_c$ & Transmit bit energy, chip duration & $\gamma_0$ & Average SNR per bit at the receiver \\
    \hline
  \end{tabular}
  }
   }
  \label{tab:notation}
\end{center}
\vspace{-6mm}
\end{table*}

\subsection{Novelty and Contributions}
Specifically, the contribution of this paper is threefold:
\begin{itemize}
\item We propose a novel DCSK-based self-sustainable RIS-aided single-input single-output (SISO) SWIPT architecture, where the RIS is partitioned into two non-overlapping surfaces. Based on the sustainability requirement, we decide on the number of RIS elements in each surface, with one acting in the EH mode to provide sufficient power to the other operating in the information transfer (IT) mode.
\item By considering the nonlinearities of the EH process and a generalized frequency-selective Nakagami-$m$ fading scenario, the harvested DC power at the RIS is characterized as a function of the transmit waveform parameters, the number of RIS elements working in the EH mode, and the channel parameters. Similarly, we characterize the associated system BER in terms of the transmit waveform parameters, the number of RIS elements functioning in the IT mode, and also the channel parameters. Moreover, for a given RIS configuration, we obtain a closed-form expression for the optimal reference length of the transmit waveform, which results in the minimum BER.
\item We demonstrate that it is not always beneficial to include more elements in the IT section of the RIS. Moreover, for a given acceptable BER, we derive a lower bound on the number of RIS elements that need to be operated in the EH mode. Further, we characterize the BER-harvested power trade-off of the proposed chaotic architecture, where we introduce the concept of an achievable `success rate-harvested power' region. Accordingly, we propose appropriate transmit waveform designs for this self-sustainable RIS-aided framework.
\end{itemize}
The derived closed-form results are verified by extensive Monte-Carlo simulations. They provide non-intuitive insights regarding how the chaotic waveform and the system parameters have an impact on the system performance. To the best of our knowledge, this is the first work that presents a complete analytical framework chaotic waveform-based waveform design for self-sustainable RIS-aided noncoherent SWIPT.

The rest of the paper is organised as follows. Section II introduces the system model and Section III presents the system performance analysis in terms of BER and average EH performance. Section IV discusses the aspect of transmit waveform design and also, it introduces the BER-harvested power trade-off. Finally, Section V presents the numerical results and Section VI concludes the paper.

\textit{Notation:} $[\cdot]^T$ denotes the transpose operator; $\mathbb{P}\{X\}, \mathbb{E}\{X\},$ and ${\rm Var}\{X\}$ represent the probability, expectation, and variance of $X$; $\Re (.)$ denotes the real part of a complex number, ${\rm erfc}(.)$ is the complementary error function, and $\Gamma (.)$ denotes the complete Gamma function. 

\section{System Model}
In this section, we provide details of the considered system model; the main mathematical notations related to the system model are summarized in Table \ref{tab:notation}.

\subsection{Network Topology}
\begin{figure*}[!t]
 \centering\includegraphics[width=0.84\linewidth]{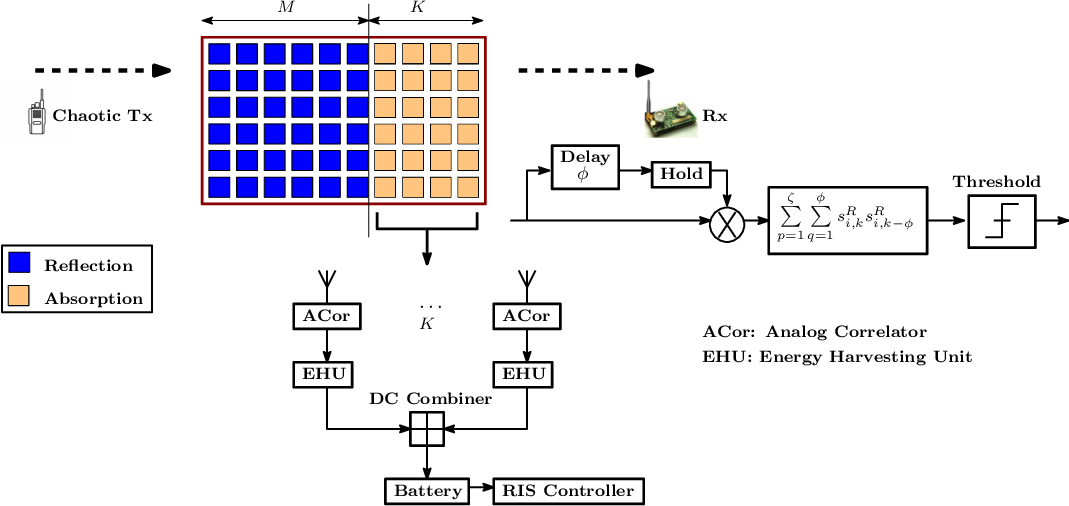}
\vspace{-2mm}
\caption{Proposed DCSK-based system architecture.}
\label{fig:model}
\vspace{-4mm}
\end{figure*}

As shown in Fig. \ref{fig:model}, we consider a SISO set-up, with a single antenna transmitter (Tx) and receiver (Rx), where the Tx employs a DCSK-based signal generator. We assume that, similar to \cite{dcskris1, dcskris2,dcskris3}, no direct link between the Tx and Rx exists due to heavy obstruction from obstacles (e.g., walls in indoor environments). Hence, an element-splitting based \cite{zeris} RIS is employed to realize the entire communication process. Specifically, the RIS consists of $N$ reflecting elements, where $M$ of them are configured for information transmission and $K$ are configured for EH, such that $M+K=N$ (discussed in Section \ref{ris}). Moreover, we consider the adjacent elements to be uncorrelated, i.e., a half-wavelength of spacing exists between them.

\subsection{Chaotic Signals}  \label{chaosdef}
\begin{figure}[!t]
 \centering\includegraphics[width=\linewidth]{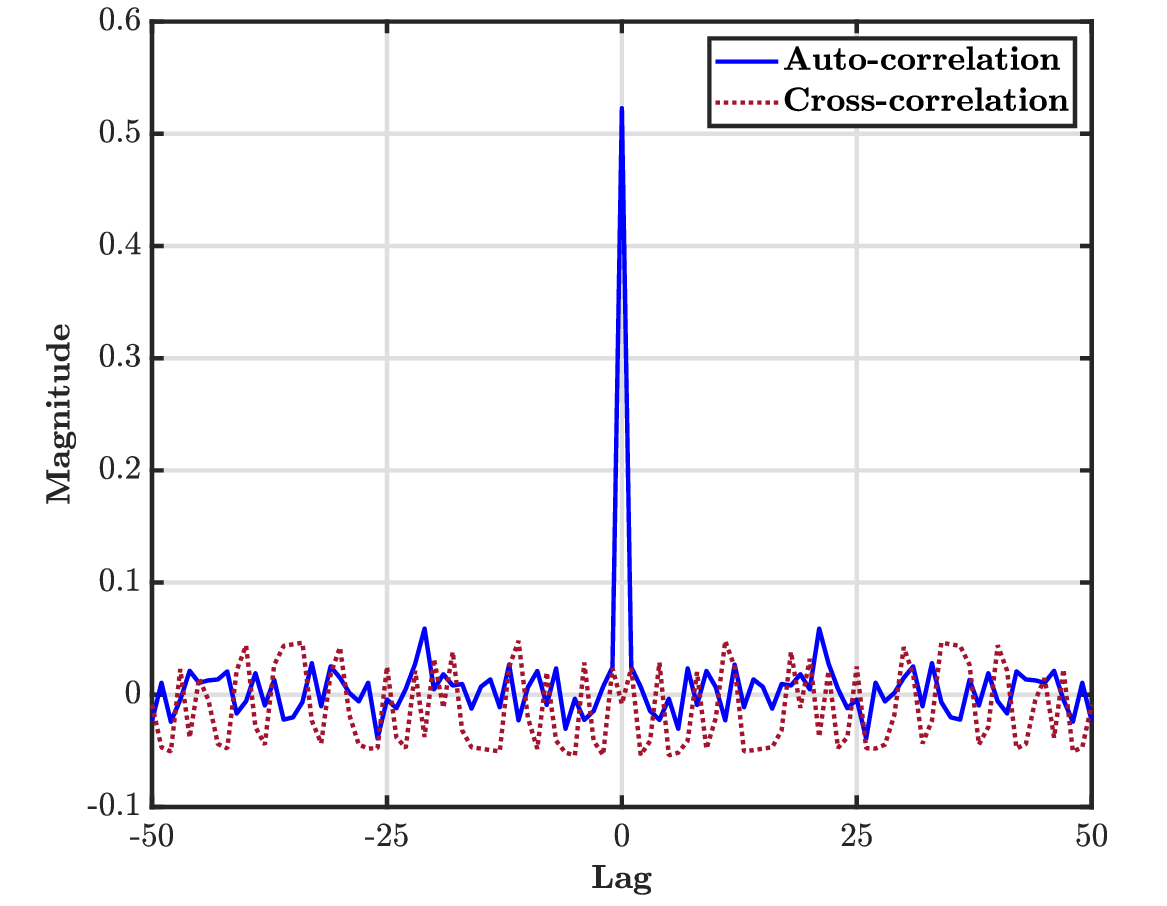}
\vspace{-6mm}
\caption{Correlation of Chebyshev map generated chaotic signals.}
\label{fig:corr}
\vspace{-6mm}
\end{figure}

Consider a DCSK signal, where a symbol is dependent on the previous one \cite{text} and different sets of chaotic sequences can be generated by using different initial conditions. Each transmitted bit is represented by two sets of chaotic signal samples, where the first set represents the reference, and the other conveys information. If the symbol $+1$ is to be transmitted, the data sample will be identical to the reference sample. Otherwise, an inverted version of the reference sample will be used as the data sample \cite{sec2}. During the $l$-th transmission interval, the transmitter output is
\begin{align}  \label{sym}
s_{l,k}=\begin{cases} 
x_{l,k}, & k=2(l-1)\beta+1,\dots,(2l-1)\beta,\\
d_lx_{l,k-\beta}, & k=(2l-1)\beta+1,\dots,2l\beta,
\end{cases}&
\end{align}
where $d_l=\pm1$ is the information symbol, $x_{l,k}$ is the chaotic sequence used as the reference signal, and $x_{l,k-\beta}$ is its delayed version. As $2\beta$ chaotic samples are being used to spread each information bit, $\beta \in \mathbb{Z}^+$ is defined as the \textit{spreading factor} \cite{sec2}. Furthermore, $x_{l,k}$ can be generated according to various existing chaotic maps. In this work, due to its good auto/cross correlation properties, we consider the Chebyshev map $x_{k+1}=4x_k^3-3x_k$ for the generation of chaotic sequences \cite{text}. As shown in Fig. \ref{fig:corr}, by `good auto/cross correlation' properties, we imply the generated chaotic signals with different initial values to be regarded as quasi-orthogonal, i.e.,
\begin{equation}  \label{corprop}
\sum_{a,b=1}^{\beta} x_a x_b \approx 0,
\end{equation}
where $a \neq b$ and $b=a+n$ for some positive integer $n$.

Unlike DCSK, in SR-DCSK, an unmodulated chaotic component of length $\phi<\beta$ is considered, followed by $\zeta$ copies of its replica, multiplied with the information, such that $\beta=\zeta\phi$ \cite{srdcsk}. Hence, the symbol duration is $\phi(1+\zeta)=\phi+\beta<2\beta$, i.e., the symbol is characterized by a sequence of $\phi+\beta$ samples of the chaotic basis signal, which is represented as
\begin{align}  \label{symsr}
& s_{l,k} \nonumber\\
 &=\!\!\begin{cases} 
x_{l,k}, & \!\!\! k=(l-1)(\beta+\phi)+1,\dots,(l-1)\beta+l\phi,\\
d_lx_{l,k-\phi}, & \!\!\!k=(l-1)\beta+l\phi+1,\dots,l(\beta+\phi).
\end{cases}&
\end{align}
\noindent Note that the SR-DCSK is a technique proposed primarily for efficient information transfer with minimum BER and also, it merges with the conventional DCSK for $\zeta=1$, i.e., $\phi=\beta$.


\subsection{RIS Characterization}  \label{ris}
An RIS consisting of $N$ reflecting elements is considered here. It is grouped into two parts, namely, IT (information transfer) and EH (energy harvesting) section, consisting of $M$ and $K$ elements respectively, with $M+K=N$. Specifically, the maximum signal-to-noise ratio (SNR) at the Rx is realized by the beam-steering functionality of the IT section, which requires a sufficient amount of energy consumption. This is realized by adjusting the phases of the reflecting elements, where the reflection coefficient for the $m$-th  element of the IT section is
\begin{equation}  \label{shiftdef}
\varphi_m=a_me^{j\theta_m} \:\: \forall \:\: m=1,\dots,M,
\end{equation}
where $a_m = 1$ and $\theta_m \in [0,2\pi)$.  The IT section is expected to intelligently reconfigure the wireless channel based on the existing CSI by varying the diagonal reflection matrix
\begin{equation}
\Phi={\rm diag} \left( \varphi_1,\cdots,\varphi_M \right).
\end{equation}
Note that for the elements belonging to the EH section, we have $\varphi_k=0$ $\forall$ $k=1,\dots,K$, i.e., the entire energy of the incoming signal is completely absorbed, which is used for the purpose of EH \cite{zeris}. This harvested energy, in turn, is employed to meet the power consumption demands of the reflecting elements as well as the RIS controller, which is responsible for providing the desired phase shifts. Consequently, by assuming that there is no stored energy at the RIS controller, the necessary amount of energy that needs to be harvested is $E_{\rm req}=T \left( MP_{\rm inf}+P_{\rm cont} \right)$, where $T$ is the time duration, $P_{\rm inf}$ denotes the power consumption of each reflecting element of the IT section, and $P_{\rm cont}$ represents the RIS controller power consumption for setting the desired phase shift of all the elements, respectively\footnote{In this work, we are considering a  passive RIS, i.e., $a_m=1$ $\forall$ $m$. Without any loss of generality, this work can be extended to an active RIS \cite{acris} scenario, where we have $a_m>1$ and $E_{\rm req}$ is a function of $a_m$} \cite{zeris}.

\subsection{Channel Model}  \label{chmodel}
We assume that the wireless links suffer from both large-scale path-loss effects and small-scale fading. Specifically, the received power at the RIS is attenuated by a factor of $C_0d_{sr}^{-\alpha_{sr}}$, where $C_0$ is the pathloss at one meter distance, $d_{sr}$ is the Tx-RIS distance, and $\alpha_{sr}$ is the corresponding path-loss exponent \cite{zeris2}. Moreover, the chaotic signals, in general, are wideband in nature \cite{2ray}, i.e., we consider a frequency-selective Tx-RIS channel with $L_{sr}$ independent paths and the received signal, at an arbitrary $n$-th element of the RIS $(n=1,\dots,N)$ is
\begin{equation}    \label{yris}
y^{\rm R}_{l,k,n}=\sqrt{P_{\rm t}C_0d_{sr}^{-\alpha_{sr}}}\sum\limits_{p=1}^{L_{sr}} h_{p,n}s_{l,k-\tau_p}+w_{sr,n},
\end{equation}
where $P_{\rm t}$ is the transmission power, $h_{p,n}$ and $s_{l,k-\tau_p}$ denote the complex channel coefficient and the delayed received signal corresponding to the $p$-th path, respectively, and $w_{sr,n}$ is the additive white Gaussian noise (AWGN) at the $n$-th RIS patch with zero mean and variance $\frac{N_0}{2}$. Furthermore, we assume that $|h_{p,n}|=\alpha_{p,n}$ is a random variable following Nakagami-$m$ fading distribution \cite{papoulis}
\begin{align}  \label{nakadefsr}
f_{\alpha_{p,n}}(\alpha)&=\frac{2m_s^{m_s}\alpha^{2m_s-1}e^{-\frac{m_s\alpha^2}{\Omega_{\alpha,p,n}}}}{\Gamma(m_s)\Omega_{\alpha,p,n}^{m_s}}, \nonumber \\
& \forall \:\:\alpha \geq 0, \:\: p=1,\dots,L_{sr} \:\: \text{and} \:\: n=1,\dots,N,
\end{align}
\noindent with $\mathbb{E}\{\alpha_{p,n}^2\}=\Omega_{\alpha,p,n}$ and $m_s \geq 1$ controls the severity of the amplitude fading. The sum of the corresponding power gains is one, i.e., $\sum\limits_{p=1}^{L_{sr}} \Omega_{\alpha,p,n}=1$ $\forall$ $n$ and identical channel statistics is considered across all the $N$ RIS elements, i.e., $\Omega_{\alpha,p,1}=\Omega_{\alpha,p,2}=\dots=\Omega_{\alpha,p,N}=\Omega_{\alpha,p}$ $\forall$ $p=1,\dots,L_{sr}$. The phase of $h_{p,n}=\theta_{h,p,n}$ is also a random variable as described in \cite{nakaang}. 
Similarly, the received power at Rx is again attenuated by $C_0d_{rd}^{-\alpha_{rd}}$, with $d_{rd}$ and $\alpha_{rd}$ being the distance and path-loss exponent corresponding to the RIS-Rx link. We model the RIS-Rx channel as a $L_{rd}$ tap frequency selective channel with complex channel coefficient $g_{n,q}$ $(n=1,\dots,N \:\: \text{and} \:\: q=1,\dots,L_{rd})$, where $|g_{n,q}|=\beta_{n,q}$ is a Nakagami-$m$ random variable
with $\mathbb{E}\{\beta_{n,q}^2\}=\Omega_{\beta,n,q}$, and $m_r \geq 1$. Here also, we have $\sum\limits_{q=1}^{L_{rd}} \Omega_{\beta,n,q}=1$ $\forall$ $n$ and identical channel statistics across all  RIS elements, i.e., $\Omega_{\beta,1,q}=\Omega_{\beta,2,q}=\dots=,\Omega_{\beta,N,q}=\Omega_{\beta,q}$ $\forall$ $q=1,\dots,L_{rd}$. Similar to the Tx-RIS channel, the phase of $g_{n,q}=\theta_{g,n,q}$ is also a random variable as stated in \cite{nakaang}.

\section{Proposed RIS-aided Chaotic System Design}
Here, we extend the SR-DCSK frame structure (Section \ref{chaosdef}) in the context of our system architecture.

\subsection{Proposed System Design}
Specifically, power transfer occurs at the RIS, where a fraction of the reflecting elements are dedicated for EH and information transfer occurs at the dedicated Rx (as seen in Fig. \ref{fig:model}). From \eqref{yris}, the received signal at the $n$-th RIS patch $(n=1,\dots,N)$ is
\begin{equation}
y^{\rm R}_{l,k,n}=\sqrt{P_{\rm t}C_0d_{sr}^{-\alpha_{sr}}}\sum\limits_{p=1}^{L_{sr}} h_{p,n}s_{l,k-\tau_p}+w_{sr,n}.
\end{equation}
Moreover, practical scenarios suggest that the largest path delay $\tau_{\max}$ is much smaller compared to the reference length $\phi$ \cite{2ray}, i.e., $0<\tau_{\max} \ll \phi$. This results in
\begin{equation}  \label{sigris}
y^{\rm R}_{l,k,n} \approx \sqrt{P_{\rm t}C_0d_{sr}^{-\alpha_{sr}}}\sum\limits_{p=1}^{L_{sr}} h_{p,n}s_{l,k}+w_{sr,n}.
\end{equation}
Accordingly, the received signal vector at the RIS is
\begin{equation}
\boldsymbol{y_{l,k}^{\rm R}}\!=\!\! \sqrt{P_{\rm t}C_0d_{sr}^{-\alpha_{sr}}}\boldsymbol{\rm H}s_{l,k}+\boldsymbol{\rm W_{sr}},
\end{equation}
where we define
\begin{align*}
\boldsymbol{y_{l,k}^{\rm R}}&=\left[ y^{\rm R}_{l,k,1}, y^{\rm R}_{l,k,2}, \cdots, y^{\rm R}_{l,k,N} \right]^T, \\
\boldsymbol{\rm H}&=\left[ \sum\limits_{p=1}^{L_{sr}} h_{p,1}, \sum\limits_{p=1}^{L_{sr}} h_{p,2}, \cdots, \sum\limits_{p=1}^{L_{sr}} h_{p,N} \right]^T, \\
\boldsymbol{\rm W_{sr}}&=\left[ w_{sr,1}, w_{sr,2}, \cdots, w_{sr,N} \right]^T.
\end{align*}
Note that, a fraction of $\boldsymbol{y_{l,k}^{\rm R}}$ is used for information transfer and the remaining portion is employed for EH. Hence, $\boldsymbol{y_{l,k}^{\rm R}}$ can be alternatively represented as
\begin{align}
\boldsymbol{y_{l,k}^{\rm R}}&=\left[ \underbrace{y^{\rm R}_{l,k,1}, \cdots, y^{\rm R}_{l,k,M}}_{\rm IT},\: \underbrace{y^{\rm R}_{l,k,M+1}, \cdots ,y^{\rm R}_{l,k,N}}_{\rm EH} \right]^T \nonumber \\
& \equiv \left[ \left( \boldsymbol{y_{l,k}^{\rm R,I}} \right)^T, \left( \boldsymbol{y_{l,k}^{\rm R,P}} \right)^T \right]^T,
\end{align}
where $\boldsymbol{y_{l,k}^{\rm R,I}}=\Big[ y^{\rm R}_{l,k,1},
\cdots, y^{\rm R}_{l,k,M} \Big]^T$ and $\boldsymbol{y_{l,k}^{\rm R,P}}=\Big[ y^{\rm R}_{l,k,M+1},
\cdots,
y^{\rm R}_{l,k,N} \Big]^T$ denote the received signal vector at the IT and EH section, respectively. Since $\boldsymbol{y_{l,k}^{\rm R,I}}$ is reflected to Rx by employing the phase shift matrix $\Phi$ at the RIS, the received signal at Rx is
\begin{align}  \label{rsig1}
y_{l,k}^{\rm Rx}& = \!\!\sqrt{C_0d_{rd}^{-\alpha_{rd}}}\boldsymbol{\rm G}^T \Phi \boldsymbol{y_{l,k}^{\rm R,I}}+w_{rd} \nonumber \nonumber \\
&\overset{(a)}{=}\delta \! \left( \! e^{j\theta_1} \sum\limits_{q=1}^{L_{rd}} g_{1,q} \sum\limits_{p=1}^{L_{sr}} h_{p,1}+e^{j\theta_2} \sum\limits_{q=1}^{L_{rd}} g_{2,q} \sum\limits_{p=1}^{L_{sr}} h_{p,2}+ \dots \right. \nonumber \\
& \left. + e^{j\theta_M} \sum\limits_{q=1}^{L_{rd}} g_{M,q} \sum\limits_{p=1}^{L_{sr}} h_{p,M}\right)s_{l,k}+w_{rd} \nonumber \\
&= \delta\left(\sum\limits_{m=1}^M e^{j\theta_m} \sum\limits_{q=1}^{L_{rd}} g_{m,q}\sum\limits_{p=1}^{L_{sr}} h_{p,m}\right)s_{l,k}+w_{rd} \nonumber \\
& \overset{(b)}{=}\delta\sum\limits_{p=1}^{L_{sr}}\sum\limits_{q=1}^{L_{rd}} \sum\limits_{m=1}^M e^{j\theta_m}h_{p,m}g_{m,q}s_{l,k}+w_{rd},
\end{align}
where we define
\begin{equation}
\boldsymbol{\rm G}=\left[
\sum\limits_{q=1}^{L_{rd}} g_{1,q}, 
\sum\limits_{q=1}^{L_{rd}} g_{2,q},
\cdots,
\sum\limits_{q=1}^{L_{rd}} g_{M,q}
\right]^T,
\end{equation}
$\delta=\sqrt{P_{\rm t}C_0^2d_{sr}^{-\alpha_{sr}}d_{rd}^{-\alpha_{rd}}},$ and $w_{rd}$ is the AWGN at the Rx with zero mean and variance $\frac{N_0}{2}$. Moreover, $(a)$ follows from the fact that the noise at the RIS, i.e., $\boldsymbol{\rm W_{sr}}$ is usually ignored in most literature as it is negligible compared to $w_{rd}$ after the reflection and going through the RIS-Rx channel. Also, we have the largest path delay $\tau_{\max}$ is much smaller compared to the reference length $\phi$ \cite{2ray} and finally, $(b)$ follows from changing the order of summation. Furthermore, since we are considering a  noncoherent modulation technique, it is not possible to have CSI at the RIS. Hence, an imperfect phase correction results in a phase error at the RIS and the received signal at the Rx is
\begin{align}  \label{rsig}
y_{l,k}^{\rm Rx}&=\delta\sum\limits_{p=1}^{L_{sr}}\sum\limits_{q=1}^{L_{rd}} \sum\limits_{m=1}^M e^{j\theta_{m}}h_{p,m}g_{m,q}s_{l,k}+w_{rd} \nonumber \\
&=\delta\sum\limits_{p=1}^{L_{sr}}\sum\limits_{q=1}^{L_{rd}} \sum\limits_{m=1}^M e^{j\theta_{e,m}}\alpha_{p,m}\beta_{m,q}s_{l,k}+w,
\end{align}
where we define the phase error $\theta_{e,m}=\theta_m+\theta_{h,p,n}+\theta_{g,n,q}$ $\forall$ $m,n,p,q$ and with a little abuse of notation, we imply $w \equiv w_{rd}$ here onwards. 

\subsection{Information and Power Transfer}
We know that $M$ out of the $N$ reflecting patches belong to the IT section of the RIS, while the remaining $N-M=K$ patches are employed to harvest energy from the incident signal. This harvested energy is used for setting the phase shift of all the $M$ patches. The Rx recovers the chaotic component from each of the transmitted frame to perform $\zeta$ partial correlations over each block of $\phi$ samples, where we have $\zeta=\frac{\beta}{\phi}$. The decision metric, corresponding to the $l$-th transmission interval, is obtained from \eqref{rsig} as expressed in \eqref{dec}.
\begin{figure*}
\begin{align}  \label{dec}
\lambda_l&= \Re \left( T_c \sum\limits_{b=1}^{\zeta}\sum\limits_{z=0}^{\phi-1} \Bigg( \delta\sum\limits_{p=1}^{L_{sr}}\sum\limits_{q=1}^{L_{rd}} \sum\limits_{m=1}^M e^{j\theta_{e,m}}\alpha_{p,m}\beta_{m,q}x_{l,z}d_l  +w_{b,z+\phi} \Bigg) \!\!\! \left( \!\delta\!\sum\limits_{p=1}^{L_{sr}}\sum\limits_{q=1}^{L_{rd}} \sum\limits_{m=1}^M e^{j\theta_{e,m}}\alpha_{p,m}\beta_{m,q}x_{l,z}+w_z \! \right)^*\! \right)\!\!. 
\end{align}
\hrule
\vspace{-4mm}
\end{figure*}
This resulting $\lambda_l$, in turn, is compared with a threshold and the actual transmitted data is recovered.

In the EH section of the RIS, we adopt a DC combining approach \cite{comb}, i.e., each reflecting element has its own energy harvesting unit (EHU), which is not affected by the remaining elements. Then, from \eqref{sigris}, the received signal at the $i$-th patch $(i=1,\dots,K)$ is given by
\begin{equation}  \label{ehsig}
y^{\rm R}_{l,k,i} = \sqrt{P_{\rm t}C_0d_{sr}^{-\alpha_{sr}}}\sum\limits_{p=1}^{L_{sr}} h_{p,i}s_{l,k},
\end{equation}
where we assume that the RIS noise is too small to be harvested. Moreover, to boost the energy harvesting performance, an analog correlator is considered prior to the EHU \cite{jstsp}. An analog correlator essentially consists of a series of delay blocks, which result in signal integration over a specified period of time; an ideal $\psi$-bit analog correlator consists of $(\psi-1)$ number of delay blocks \cite{anaco2}. Therefore, as the transmit symbol duration is $\phi+\beta$, we consider $\psi=\phi+\beta$. Accordingly, based on \eqref{ehsig}, the analog correlator output, at the $i$-th patch, with respect to the $l$-th transmitted symbol, is given by
\begin{align}  \label{corr}
y^{\rm AC}_{l,i}&=\sum\limits_{k=1}^{\phi+\beta} y^{\rm R}_{l,k,i}=\sqrt{P_{\rm t}C_0d_{sr}^{-\alpha_{sr}}}\sum\limits_{k=1}^{\phi+\beta}\sum\limits_{p=1}^{L_{sr}} h_{p,i}s_{l,k} \nonumber \\
&=\sqrt{P_{\rm t}C_0d_{sr}^{-\alpha_{sr}}} \left( \sum\limits_{p=1}^{L_{sr}} h_{p,i} \right) \left( \sum\limits_{k=1}^{\phi+\beta}s_{l,k} \right).
\end{align}
This resulting signal acts as input to the EHU, which basically consists of a diode followed by a passive low pass filter, i.e., it acts as an envelope
detector \cite{comb}. The output DC power from each of the reflecting patches are combined together by a DC combining circuit \cite{comb}. Based on the nonlinearity of the EHUs, the total output DC power is expressed as $P_{\rm harv}=\sum\limits_{i=1}^K\frac{v_{i, {\rm out}}^2}{R_L}$, where $R_L$ is the load resistance and the output DC voltage of the EHU at the $i$-th patch, i.e., $v_{i, {\rm out}}$ is expressed as \cite{comb}
\begin{equation}  \label{brunoeh}
v_{i, {\rm out}}=\nu_1\mathbb{E} \{ |y^{\rm AC}_{l,i}|^2 \}+\nu_2\mathbb{E} \{ |y^{\rm AC}_{l,i}|^4 \}.
\end{equation}
Here, the parameters $\nu_1$ and $\nu_2$ are constants determined by the characteristics of the circuit. For the sake of presentation, we will use $\mu_1=\nu_1P_{\rm t}C_0d_{sr}^{-\alpha_{sr}}$ and $\mu_2=\nu_2 \left( P_{\rm t}C_0d_{sr}^{-\alpha_{sr}}\right)^2$. Thereafter, by combining \eqref{corr} and \eqref{brunoeh}, we obtain
\begin{align}
v_{i, {\rm out}}& = \mu_1 \mathbb{E} \left\lbrace \Bigg | \left( \sum\limits_{p=1}^{L_{sr}} h_{p,i} \right) \left( \sum\limits_{k=1}^{\phi+\beta}s_{l,k} \right) \Bigg |^2 \right\rbrace \nonumber \\
& + \mu_2 \mathbb{E} \left\lbrace \Bigg | \left( \sum\limits_{p=1}^{L_{sr}} h_{p,i} \right) \left( \sum\limits_{k=1}^{\phi+\beta}s_{l,k} \right) \Bigg |^4 \right\rbrace
\end{align}
Here, the expectation is taken over both the channel and the transmit waveform. Moreover, we know that $|h_{p,i}|=\alpha_{p,i}$ (defined in Section II-D) and also, if two arbitrary random variables, say $X$ and $Y$ are independent, then for any functions $a$ and $b$, we have 
\begin{equation}
\mathbb{E} \{ a(X)b(Y)\}=\mathbb{E} \{ a(X)\}\mathbb{E} \{b(Y)\}.
\end{equation}
As a result, in this case, we obtain
\begin{align}
v_{i, {\rm out}}&=\mu_1\mathbb{E} \left\lbrace \left( \sum\limits_{p=1}^{L_{sr}} \alpha_{p,i} \right)^2\right\rbrace \mathbb{E} \left\lbrace \left( \sum\limits_{k=1}^{\phi+\beta}s_{l,k} \right)^2 \right\rbrace \nonumber \\
& +\mu_2 \mathbb{E} \left\lbrace \left( \sum\limits_{p=1}^{L_{sr}} \alpha_{p,i} \right)^4\right\rbrace \mathbb{E} \left\lbrace \left( \sum\limits_{k=1}^{\phi+\beta}s_{l,k} \right)^4 \right\rbrace \nonumber \\
&= \mu_1\mathbb{E}\{ \mathcal{H}_i^2 \} \mathbb{E} \{ \mathcal{S}^2 \} + \mu_2\mathbb{E}\{ \mathcal{H}_i^4 \} \mathbb{E} \{ \mathcal{S}^4 \},
\end{align}
where we define 
\begin{equation*}
\mathcal{H}_i=\sum\limits_{p=1}^{L_{sr}} \alpha_{p,i} \:\: \text{and} \:\: \mathcal{S}=\sum\limits_{k=1}^{\phi+\beta}s_{l,k}.
\end{equation*}
Accordingly, the harvested DC power is
\begin{equation}  \label{pharv}
P_{\rm harv}=\frac{1}{R_L} \sum\limits_{i=1}^K v_{i, {\rm out}}^2,
\end{equation}
where we assume identical load resistance across all the $K$ reflecting patches\footnote{Note that, we have adopted a DC combining approach. Similar to the receiver design proposed in \cite{twc}, an RF combining approach may also be employed, where the received signal at all the reflecting patches are combined together and then fed to the EHU.}.

\section{Waveform Design Investigating Reflect-Harvest Trade-off}
In this section, we first analyse the performance of the proposed RIS-aided DCSK-based system architecture; specifically, we investigate the aspect of average BER and EH performance. Thereafter, based on the application specific requirements of acceptable BER or harvested DC power or both, we investigate and characterize the reflect/harvest trade-off at the RIS to decide on the best $(M,K)$ combination.

\subsection{BER Performance}
Here we investigate the impact of SR-DCSK signals on the information receiver of the proposed system. If $T_c$ is the chip duration, based on the SR-DCSK frame structure \eqref{symsr}, the corresponding transmitted bit energy is defined as
\begin{equation}  \label{biten}
E_{\rm b}=P_{\rm t}T_c(\beta+\phi)\mathbb{E}\left\lbrace x_k^2 \right\rbrace,
\end{equation}
where $x_k$ is the chaotic chip as defined in Section \ref{chaosdef}. Accordingly, we define 
\begin{equation}  \label{gamdef}
\gamma_0=\frac{E_{\rm b}C_0^2 d_{sr}^{-\alpha_{sr}}d_{rd}^{-\alpha_{rd}}}{N_0}
\end{equation}
and by using \eqref{dec} and considering $T_c=1$ for mathematical simplicity, we obtain the decision variable $\lambda_l$ corresponding to the $l$-th transmitted symbol. Now, based on this quantity $\lambda_l$, we analyze the BER performance by the following theorem.
\begin{theorem}  \label{theo1}
If $M(\leq N)$ RIS elements are employed in the IT section, the system's BER is obtained as
\begin{equation}  \label{bertheo}
{\rm BER}=\frac{1}{2}\int\limits_0^{\infty} {\rm erfc} \left( \left[ \frac{\left( 1+\zeta \right)^2}{\gamma_0\zeta\Lambda} + \frac{\left( \beta+\phi \right)^2}{2\beta\gamma_0^2\Lambda^2}\right]^{-\frac{1}{2}}   \right)f(\Lambda) d\Lambda,
\end{equation}
where we define $\Lambda=\sum\limits_{p=1}^{L_{sr}}\sum\limits_{q=1}^{L_{rd}} \Big|\sum\limits_{m=1}^M e^{j\theta_{e,m}}\alpha_{p,m}\beta_{m,q}\Big|^2$ and $f(\Lambda)$ is its probability distribution function.
\end{theorem}
\begin{proof}
See Appendix \ref{app1}.
\end{proof}
We observe from Theorem \ref{theo1}, that the BER performance is a function of the transmit waveform parameters, i.e., $\beta$ and $\phi$, number of RIS patches that are being utilized in the IT mode, i.e., $M$, and finally, on the wireless channel parameters, i.e., $m_s,m_r,\Omega_{\alpha,p}$ $\forall$ $p=1,\dots,L_{sr}$, $\Omega_{\beta,q}$ $\forall$ $q=1,\dots,L_{rd}$. However, we observe that it is difficult to obtain an analytical closed-form solution in \eqref{bertheo}. As a result, similar to the approach in \cite{srdcsk}, we investigate an AWGN channel scenario in this context. Next, we comment on the optimality of $\phi$, for a given set of system parameters.

\begin{theorem}  \label{theo2}
In the AWGN scenario, for a given set of system parameters, i.e., $\beta$ and $\gamma_0$, and $M$, the BER is evaluated as
\begin{equation}
{\rm BER}=\frac{1}{2} {\rm erfc} \left(  \left[ \frac{\left( 1+\zeta \right)^2}{\gamma_0\zeta \Lambda} + \frac{\left( \beta+\phi \right)^2}{2\beta\gamma_0^2\Lambda^2}\right]^{-\frac{1}{2}}   \right),
\end{equation}
where we define $\Lambda=\sum\limits_{i=1}^M\sum\limits_{j=1}^M \cos \left( \theta_{e,i}-\theta_{e,j} \right)=M+ 2\sum\limits_{\substack{i,j=1 \\ i \neq j}}^M \cos \left( \theta_{e,i}-\theta_{e,j} \right)$. Accordingly, the corresponding optimal reference length with respect to having minimum BER is
\begin{equation}
\phi_{\min}=\frac{\gamma_0\Lambda}{2} \left( \sqrt{1+\frac{4\beta}{\gamma_0\Lambda}}-1 \right).
\end{equation}
\end{theorem}
\begin{proof}
See Appendix \ref{app2}.
\end{proof}
Note that, we obtain $\Lambda= M + 2 \sum\limits_{i=1}^{M-1} i = M^2$ for $\theta_{e,i}=\theta_{e,j}$ $\forall$ $i,j=1,\dots,M$. Also, the special case of $\theta_{e,i}=\theta_{e,j} \pm \frac{\pi}{2}$ $\forall$ $i \neq j$ and $i,j=1,\dots,M$ results in  $\Lambda=M$. Hence, it is interesting to observe that in case of complete CSI at the RIS, we have $\theta_{e,i}=\theta_{e,j}=0$ $\forall$ $i,j=1,\dots,M$, i.e., $\Lambda=M^2$. This, in turn, demonstrates the importance of CSI and its direct impact on the BER characterization of the system. Moreover, apart from the system parameters, i.e., $\beta$ and $\gamma_0$, we also observe, that both BER and $\phi_{\min}$ are functions of the channel statistics as well as $M$, i.e., the number of reflecting elements in the IT section of the RIS. Furthermore, we know from Section \ref{chaosdef} that $\beta=\zeta\phi$, where the number of replicas $\zeta \in \mathbb{Z}^+$. As a result, in scenarios where $\frac{\beta}{\phi_{\min}} \notin \mathbb{Z}^+$, we choose an appropriate $\phi \in \mathbb{Z}^+$ that is closest to $\phi_{\min}$.
\begin{corr}  \label{corr1}
For the scenario of $\gamma_0\Lambda \gg 4\beta$, classical DCSK results in the best BER performance.
\end{corr}
The above corollary follows directly from the Taylor series expansion of $\sqrt{1+\frac{4\beta}{\gamma_0\Lambda}}$ when $\gamma_0\Lambda \gg 4\beta$ holds, i.e., we have
\begin{equation}
\phi_{\min}=\frac{\gamma_0\Lambda}{2} \left( \sqrt{1+\frac{4\beta}{\gamma_0\Lambda}}-1 \right) \approx \beta.
\end{equation}
This is an interesting observation, which illustrates, that for a specific scenario, the classical DCSK outperforms SR-DCSK in terms of BER performance. In other words, always employing more and more reflecting elements in the IT section of the RIS is not beneficial. Finally, Theorem \ref{theo2} demonstrates that for a given set of system parameters and channel statistics, the BER illustrates an unimodal variation against $\phi$ with the minimum being attained at $\phi=\phi_{\min}$, beyond which it starts to  increase when $\phi \in (\phi_{\min},\beta]$. Thus, with our objective being minimization of BER, we restrict ourselves in $\phi \in [1,\phi_{\min}]$ and later in this work, unless specified, we are always interested in this region of operation.

\subsection{Average EH Performance}
Here we investigate the energy harvesting performance at the EH section of the RIS, which consists of $K(\leq N)$ reflecting elements. By neglecting the contribution of the noise component, in the following theorem, we analyze the harvested DC power accordingly.

\begin{theorem}  \label{theo3}
If the EH section of the RIS consists of $K$ reflecting elements, the harvested power is obtained as
\begin{equation}  \label{pharvfinal}
P_{\rm harv}=\frac{K\Upsilon}{R_L},
\end{equation}
where $\Upsilon$ is defined as \eqref{upsidef}.

\begin{figure*}[t]
\begin{align}  \label{upsidef}
&\Upsilon=\left( \mu_1\phi\left(  1+\zeta^2 \right) \left( \frac{1}{2}+\frac{1}{m_s}\left(\frac{\Gamma(m_s+0.5)}{\Gamma(m_s)}\right)^2 \sum_{\mathclap{\substack{p_1,p_2=1\\p_1 \neq p_2}}}^{L_{sr}} \sqrt{\Omega_{\alpha,p_1}\Omega_{\alpha,p_2}} \right) \right. \nonumber \\
& \left. + 9\mu_2 \phi\left( 1+6\zeta^2+\zeta^4 \right) \left( 2\phi-1 \right) \!\! \left(\sum\limits_{k_1+k_2+\dots+k_{L_{sr}}=4}\! \frac{1}{k_1! \: k_2! \: \dots \: k_{L_{sr}}!}\! \prod\limits_{p=1}^{L_{sr}}  \frac{\Gamma(m_s+\frac{k_p}{2})}{\Gamma(m_s)}\left(\frac{\Omega_{\alpha,p}}{m_s}\right)^{\frac{k_p}{2}} \!\right)\!\! \right)^2\!\!.
\end{align}
\hrule
\vspace{-4mm}
\end{figure*}
\end{theorem}
\begin{proof}
See Appendix \ref{app3}
\end{proof}
Theorem \ref{theo3} characterizes the harvested DC power at the EH section of the RIS in terms of the transmit waveform parameters, the channel parameters, and also the number of reflecting elements in the EH section. Note that \eqref{pharvfinal} is a lengthy and complex equation. Hence, to obtain analytical insights, we consider a flat fading scenario in the following corollary.

\begin{corr}
For flat fading, with $K$ reflecting elements in the EH section of the RIS, we obtain
\begin{align}  \label{pharvflat}
&P_{\rm harv}=\frac{K}{R_L} \left(  \frac{\mu_1\phi\left(  1+\zeta^2 \right)}{2}  \right. \nonumber \\
& \left. + \frac{3}{8}\mu_2 \phi\left( 1+6\zeta^2+\zeta^4 \right)\left( 2\phi-1 \right) \left( \frac{m_s+1}{m_s} \right)\right)^2\!\!.
\end{align}
\end{corr}
The above corollary follows directly from using $L_{sr}=1$ and $\frac{\Gamma(m_s+2)}{m_s^2\Gamma(m_s)}=\frac{m_s+1}{m_s}$. From \eqref{pharvflat}, for a given set of system parameters, we observe the impact of $K$ on the harvested DC power. Specifically, $P_{\rm harv}$ increases monotonically with $K$, which appears to be similar to the DC combining approach proposed in \cite{comb}. Moreover, we observe that $P_{\rm harv}$ decreases with increasing $m_s$, i.e., fading is beneficial for energy harvesting, which supports the claims made in \cite{fhelps} regarding the benefits of fading on the EH process. Furthermore, we also note from \eqref{pharvflat} that with all other parameters remaining constant, the maximum harvesting performance is obtained by using $\phi=1$ and $\zeta=\beta$. This claim corroborates the observations made in \cite{jstsp} regarding WPT optimal DCSK-based chaotic waveforms.

\subsection{Reflect-Harvest Trade-off Characterization}
Our primary objective is to guarantee that the energy harvested at the EH section of the RIS must be sufficient to meet the the energy requirement at the RIS controller for the purpose of providing phase shift at all the $M$ reflecting elements, which are being used in the IT mode (with a satisfactory BER performance). In this context, there are two aspects to the problem.
\begin{enumerate}
\item For fixed $\beta,\zeta,$ and $\phi$, we investigate the best possible $(M,K)$ combination.
\item For fixed $(M,K)$ combination, we look into the optimal transmit waveform design
\end{enumerate}
The non-triviality of this problem lies in the fact that energy and information contents are contrasting ideas as far as the issue of transmit waveform design is concerned. As observed in the previous section, optimal waveform design with minimum BER leads to different values of $\phi$ whereas for a given $\beta$, the maximum harvested power is obtained at $\phi=1$. Moreover, the specific $\phi$ corresponding to the minimum BER is also a function of $M$, which in turn affects the EH performance at the RIS, as $K=N-M$.

\subsection*{C-I. Best $(M,K)$ Combination for Fixed $\beta,\zeta,$ and $\phi$}
If the acceptable BER at the Rx is ${\rm BER}_0$, we observe from \eqref{bertheo} that for a given set of system parameters, ${\rm BER}_0$ is a function of $M$. Hence,  as stated in Section \ref{ris}, the necessary amount of energy that needs to be harvested in the EH section of the RIS is $E_{\rm req}=T \left( MP_{\rm inf}+P_{\rm cont} \right)$. Therefore, from \eqref{pharvfinal}, we obtain the following lower bound on $K$, i.e.,
\begin{align}  \label{kmindef}
K \geq K_{\min}=\frac{R_L E_{\rm req}}{\Upsilon},
\end{align}
where $\Upsilon$ is already defined in \eqref{upsidef}. As a result, we have $K \in [K_{\min},N-M]$. However, it is to be noted that, we have implicitly assumed $K_{\min} \leq N-M$, which may not always hold. Such scenarios lead to what may be defined as \textit{energy outage}, which is not desired for any application. As a result, in such cases, we investigate the aspect of transmit waveform design for a fixed $(M,K)$ pair.

\subsection*{C-II. Transmit Waveform Design for Fixed $(M,K)$}
Theorem \ref{theo2} demonstrates that for a given set of system and channel parameters, $\phi_{\min}$ is a function of $M$.  Accordingly, we evaluate $\phi_{\min}$ and the associated ${\rm BER}_{\min}$ for a given $M$. Moreover, for a fixed $\beta$, due to the unimodal variation of BER against $\phi$ as observed in Theorem \ref{theo2}, the application-specific acceptable ${\rm BER}_0$ results in two values of $\phi$, namely $\phi_A$ and $\phi_B$. On the contrary, from Theorem \ref{theo3} we observe, that for a given set of parameters and $K=N-M$, the maximum harvested DC power is obtained for $\phi=1$ and it monotonically decreases with increasing $\phi$. This implies, that to meet a certain energy budget, more number of reflecting elements need to be in the EH section of the RIS, i.e., higher value of $K$ is required if we choose a higher value of $\phi$.

But, in order to attain ${\rm BER}_0$, we have $\phi \in [\phi_A,\phi_B]$, where $\phi_A \leq \phi_{\min} \leq \phi_B$. However, beyond $\phi=\phi_{\min}$, the BER starts increasing again and as a result, we restrict ourselves to $[\phi_A,\phi_{\min}]$. Accordingly, for a fixed $(M,K)$ pair, by assuming that the Tx transmits SR-DCSK frames of length $\beta+\phi$ and defining a quantity `success rate' ${\rm SR}=1-{\rm BER}$ as proposed in \cite{twc}, we propose the ${\rm SR}-P_{\rm harv}$ region definition as stated in \eqref{regiondef}.
\begin{figure*}[t]
\begin{align}  \label{regiondef}
&\mathcal{C}_{SR-P_{\rm harv}}\!\left( \phi : \phi \in [\phi_A,\phi_{\min}]\right)= \! \left\lbrace \! \!\left( SR,P_{\rm harv} \right):\! \frac{1}{2} {\rm erfc} \left(  \left[ \frac{\left( 1+\zeta \right)^2}{\gamma_0\zeta \Lambda} + \frac{\left( \beta+\phi \right)^2}{2\beta\gamma_0^2\Lambda^2}\right]^{-\frac{1}{2}}   \right) \leq 1-{\rm SR}_0, P_{\rm harv} \geq MP_{\rm inf}+P_{\rm cont} \right\rbrace .
\end{align}
\vspace{-2mm}
\hrule
\vspace{-2mm}
\end{figure*}
Here, ${\rm SR}_0=1-{\rm BER}_0$ defines the minimum acceptable success rate. Moreover, as $(M,K)$ is fixed in this characterization, $\phi$ is the only deciding factor here. Note that, for a given $\beta$, we have $\beta=\zeta\phi$, where $\zeta$ is the number of replicas of the modulated chaotic component of length $\phi$. Hence, the chaotic repetition mechanism forms an integral part of the proposed ${\rm SR}-P_{\rm harv}$ region. We observe, that the novelty here lies in the transmit waveform design, which guarantees that ${\rm BER}_0$ will be attained by the $M$ reflecting elements and the energy required for this performance will be harvested by the remaining, i.e., $K=N-M$ RIS elements. Furthermore, we also demonstrate in Theorem \ref{theo2} that $\phi_{\min}$ is a function of $M$, unlike the work in \cite{twc}. Hence, we can say, that the proposed ${\rm SR}-P_{\rm harv}$ region definition for this RIS-aided system architecture depends on the parameters of the transmit waveform, the wireless channel, and also on the choice of $(M,K)$ pair, such that $M+K=N$, a constant.

Note that from the perspective of IT, the Rx of the proposed framework corresponds to the conventional SR-DCSK receiver \cite{srdcsk}. Secondly, due to the inherent noncoherent nature of DCSK-based systems, it avoids the CSI acquisition cost completely \cite{text}. Thirdly, based on the channel statistics and system parameters, the computation regarding the number of RIS elements allocated to the IT and EH processes, respectively is decided a-priori. Therefore, this process is not `real-time' but based on the statistical channel knowledge, i.e., it is a one-time calculation and hence, the associated computational complexity is significantly low.

\subsection{Comparison with DCSK}
Finally, we characterize the performance gain of the proposed transmit waveform design over its existing conventional DCSK-based counterpart. Specifically, for a given $\beta,\gamma_0$ and $(M,K)$ pair, we denote the $\{{\rm success \:\: rate}, {\rm harvested \:\: power}  \}$ pair of the proposed waveform design and DCSK as $\left\lbrace {\rm SR_p},P_{\rm harv,p} \right\rbrace $ and $\left\lbrace {\rm SR_d},P_{\rm harv,d} \right\rbrace $, respectively. Accordingly, from Theorem \ref{theo2}, we obtain
\begin{equation}
{\rm SR_p}=1-\frac{1}{2} {\rm erfc} \left(  \frac{1}{\sqrt{\Psi(\phi)}}   \right),
\end{equation}
where $\Psi(\phi)=\frac{\left( 1+\zeta \right)^2}{\gamma_0\zeta \Lambda} + \frac{\left( \beta+\phi \right)^2}{2\beta\gamma_0^2\Lambda^2}$. Therefore, the maximum SR is attained for $\phi_{\rm opt}=\frac{\gamma_0\Lambda}{2} \left( \sqrt{1+\frac{4\beta}{\gamma_0\Lambda}}-1 \right)$. On the contrary, for the DCSK-based counterpart, we replace $\phi=\beta$ to obtain
\begin{equation}
{\rm SR_d}=1-\frac{1}{2} {\rm erfc} \left(  \frac{1}{\sqrt{\Psi(\beta)}}   \right).
\end{equation}
Based on the properties of the ${\rm erfc}(\cdot)$ function and the fact that $\phi \leq \beta$, we obtain
\begin{equation}  \label{srcomp}
{\rm SR_d} \leq {\rm SR_p}.
\end{equation}
Now, in both the cases, we obtain the harvested power from Theorem \ref{theo3} as
\begin{align}  \label{hp}
P_{\rm harv,p}&=\frac{K}{R_L} \left( \mu_1\chi_1\phi \left( 1+\zeta^2 \right)\right. \nonumber \\ &+\left.9\mu_2\chi_2\phi\left( 1+6\zeta^2+\zeta^4 \right) \left( 2\phi-1 \right) \right)^2
\end{align}
and
\begin{align} \label{hd}
P_{\rm harv,d}&=\frac{K}{R_L} \left( 2\mu_1\chi_1\beta + 72\mu_2\chi_2\beta \left( 2\beta-1 \right) \right)^2,
\end{align}
where we define
\begin{equation*}
\chi_1=\frac{1}{2}+\frac{1}{m_s}\left(\frac{\Gamma(m_s+0.5)}{\Gamma(m_s)}\right)^2 \sum_{\mathclap{\substack{p_1,p_2=1\\p_1 \neq p_2}}}^{L_{sr}} \sqrt{\Omega_{\alpha,p_1}\Omega_{\alpha,p_2}}
\end{equation*}
and
\begin{align*}
\chi_2&=\sum\limits_{k_1+k_2+\dots+k_{L_{sr}}=4}\! \frac{1}{k_1! \: k_2! \: \dots \: k_{L_{sr}}!}\! \nonumber \\
& \qquad\qquad\times \prod\limits_{p=1}^{L_{sr}}  \frac{\Gamma(m_s+\frac{k_p}{2})}{\Gamma(m_s)}\left(\frac{\Omega_{\alpha,p}}{m_s}\right)^{\frac{k_p}{2}}.
\end{align*}
As $\beta \geq \phi$, by comparing \eqref{hp} and \eqref{hd}, we obtain
\begin{equation}  \label{phcomp}
P_{\rm harv,d} \leq P_{\rm harv,p}.
\end{equation}
Hence, by combining \eqref{srcomp} and \eqref{phcomp}, we conclude that the proposed waveform design results in an enhanced ${\rm SR}-P_{\rm harv}$ region as compared with its DCSK-based counterpart.

\section{Numerical Results}
In this section, we evaluate the considered system model to validate our proposed theoretical analysis. Unless otherwise stated, we consider a transmission power of $P_{\rm t}=30$ dBm, pathloss at one meter distance $C_0=10^{-3.53}$, and power consumption of the controller and each element of the IT section of the RIS being $P_{\rm cont}=50$ mW and $P_{\rm inf}=2$ $\mu$W, respectively \cite{zeris}. The parameters considered for the non-linear EH model are: $\nu_1=0.9207 \times 10^3, \nu_2=0.0052 \times 10^9,$ and $R_L=5000$ $\Omega$ \cite{comb}. Without any loss of generality, we consider a two-tap frequency selective Nakagami-$m$ fading scenario for both the Tx-RIS and RIS-Rx channel with the fading parameter $m_s=m_r=m$ and the path-loss exponent $\alpha_{sr}=\alpha_{rd}=3$. As the considered set of parameter values is used for the sake of presentation, different sets of values will affect the performance but will lead to similar observations.

\begin{figure}[!t]
 \centering\includegraphics[width=\linewidth]{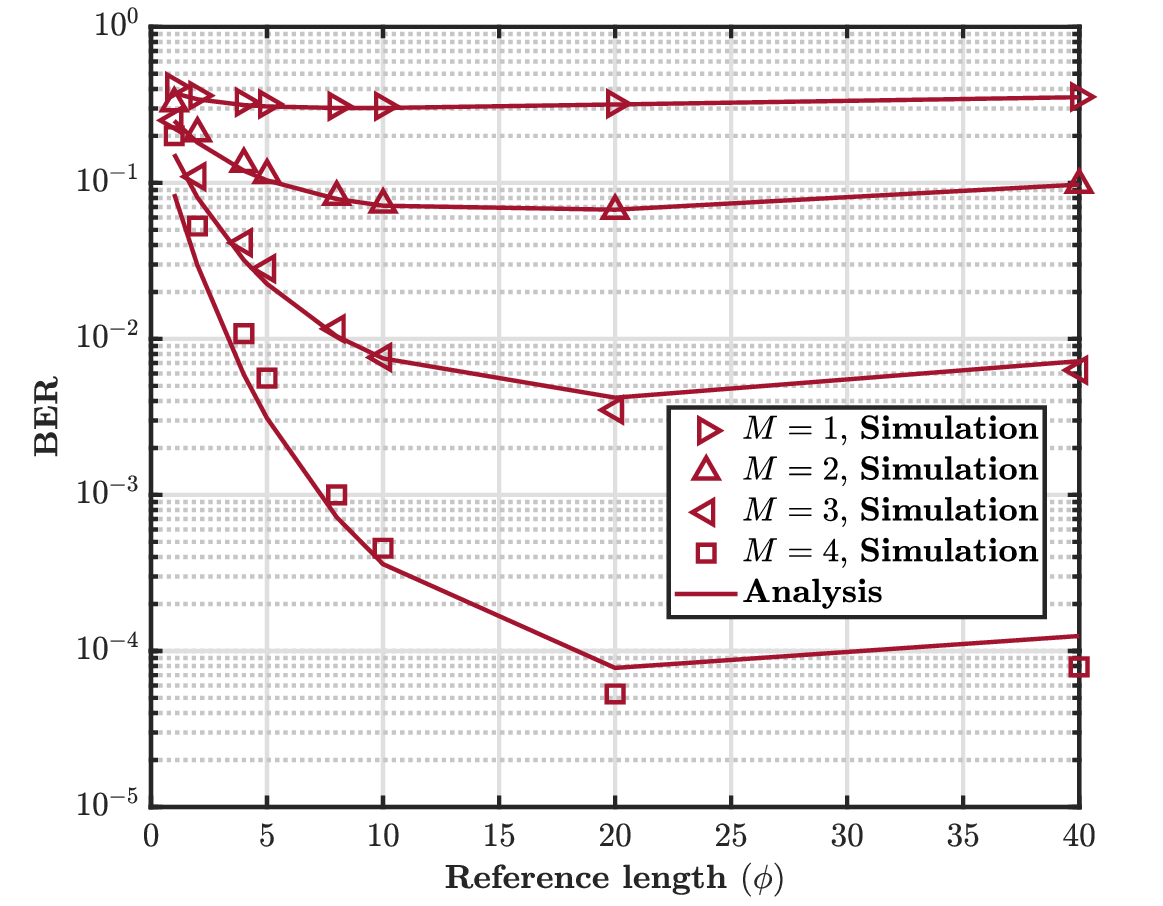}
\vspace{-2mm}
\caption{Impact of $M$ on BER; $\gamma_0=4$ dB and $\beta=40$.}
\label{fig:res1}
\vspace{-2mm}
\end{figure}

For various $M$, Fig. \ref{fig:res1} illustrates the BER achieved by the considered system model, in an AWGN scenario, where lines and markers correspond to analytical and simulation results, respectively. By assuming no phase error at the RIS and considering $\beta=40$ and $\gamma_0=4$ dB, we investigate the impact of $M$ on BER as a function of the reference length $\phi$. We observe that the optimal reference length $\phi_{\min}$ is different for different values of $M$. Specifically, we have $\phi_{\min}=8.84,15.65,20.82,24.75$ for $M=1,2,3,4$, respectively (Theorem \ref{theo2}) and as $\frac{\beta}{\phi} \in \mathbb{Z}^+$, we have $\phi_{\min}=8$ for $M=1$ and $\phi_{\min}=20$ for $M=2,3,4$, respectively. It is interesting to note, that $M$ does impact $\phi_{\min}$; increasing $M$ also results in rapidly decreasing BER with $\phi$. Moreover, for a fixed $M$, BER initially decreases with $\phi$ before attaing its minimum value at $\phi=\phi_{\min}$ and then, it starts increasing again. Finally, irrespective of $M$, there is a gap between the analytical and simulation results for lower $\phi$, which decreases as $\phi$ increases. This is due to the considered Gaussian approximation in the theoretical derivation of BER (Appendix \ref{app1}).

\begin{figure}[!t]
 \centering\includegraphics[width=\linewidth]{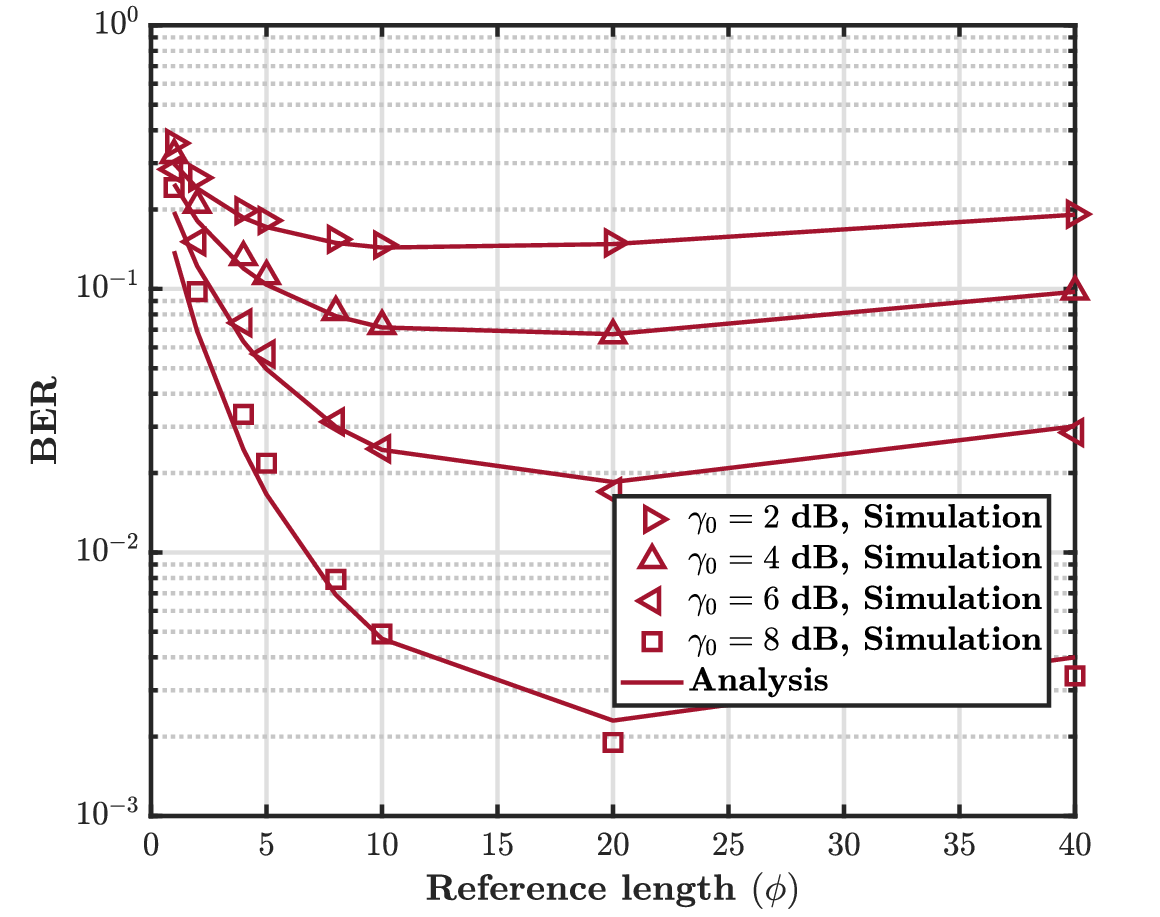}
\vspace{-2mm}
\caption{Impact of $\gamma_0$ on BER; $M=2$ and $\beta=40$.}
\label{fig:res2}
\vspace{-2mm}
\end{figure}

By considering $\beta=40$ and $M=2$ in an  identical set-up as in Fig. \ref{fig:res1}, Fig. \ref{fig:res2} demonstrates the impact of $\gamma_0$ on BER as a function of the reference length $\phi$. We observe that, for a given $M$, unequal values of $\gamma_0$ result in different $\phi_{\min}$. Moreover, the figure depicts that, for a fixed $\gamma_0$, the unimodal nature of the variation of BER with $\phi$ confirms the analytical derivation in Theorem \ref{theo2}. Furthermore, we also observe that the rate of variation of BER with $\phi$ increases with higher $\gamma_0$. Finally, the reason behind the difference in the analytical and simulation results for lower values of $\phi$ is already explained above.

\begin{figure}[!t]
 \centering\includegraphics[width=\linewidth]{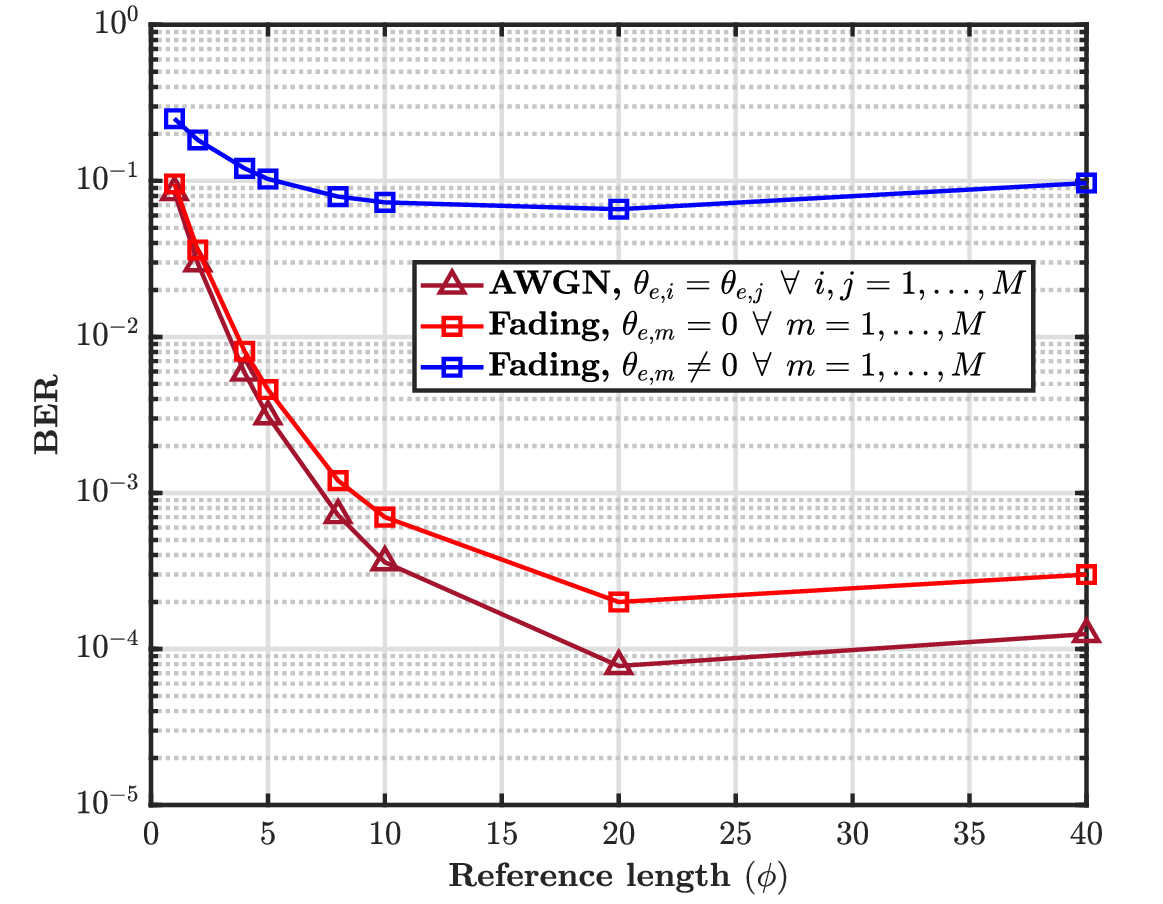}
\vspace{-2mm}
\caption{Impact of channel on BER; $M=4,\gamma_0=4$ dB and $\beta=40$.}
\label{fig:berch}
\vspace{-2mm}
\end{figure}

Fig. \ref{fig:berch} shows the impact of the wireless channel on BER achieved by the considered system model. In this figure, with $M=4,\gamma_0=4$ dB, and $\beta=40$, we consider the following  simulation environments.
\begin{enumerate}
\item AWGN channel with no phase error at the RIS, i.e., $\theta_{e,i}=\theta_{e,j}$ $\forall$ $i,j=1,\dots,M$.
\item Nakagami-$m$ fading scenario with $m=4,$ $\Omega_{\alpha,1}=\Omega_{\beta,1}=0.8,$ and $\Omega_{\alpha,2}=\Omega_{\beta,2}=0.2$. Here, both the possibilities of CSI availability, and unavailability at the RIS are taken into account, i.e., $\theta_{e,m}=0$ and $\theta_{e,m} \neq 0$ $\forall$ $m=1,\dots,M$, respectively.
\end{enumerate}
We observe from the figure that, with all the system parameters remaining constant, the considered AWGN scenario provides the lower bound of the system BER performance. Moreover, the figure also demonstrates the impact of having (and not having) CSI at the RIS. However, since we employ a noncoherent modulation technique, practically it is not possible to have CSI at the RIS, i.e., $\theta_{e,m} \neq 0$ $\forall$ $m$. Furthermore, note that, if it is possible to obtain CSI at the RIS, it significantly enhances the BER performance of the system.

\begin{figure}[!t]
 \centering\includegraphics[width=\linewidth]{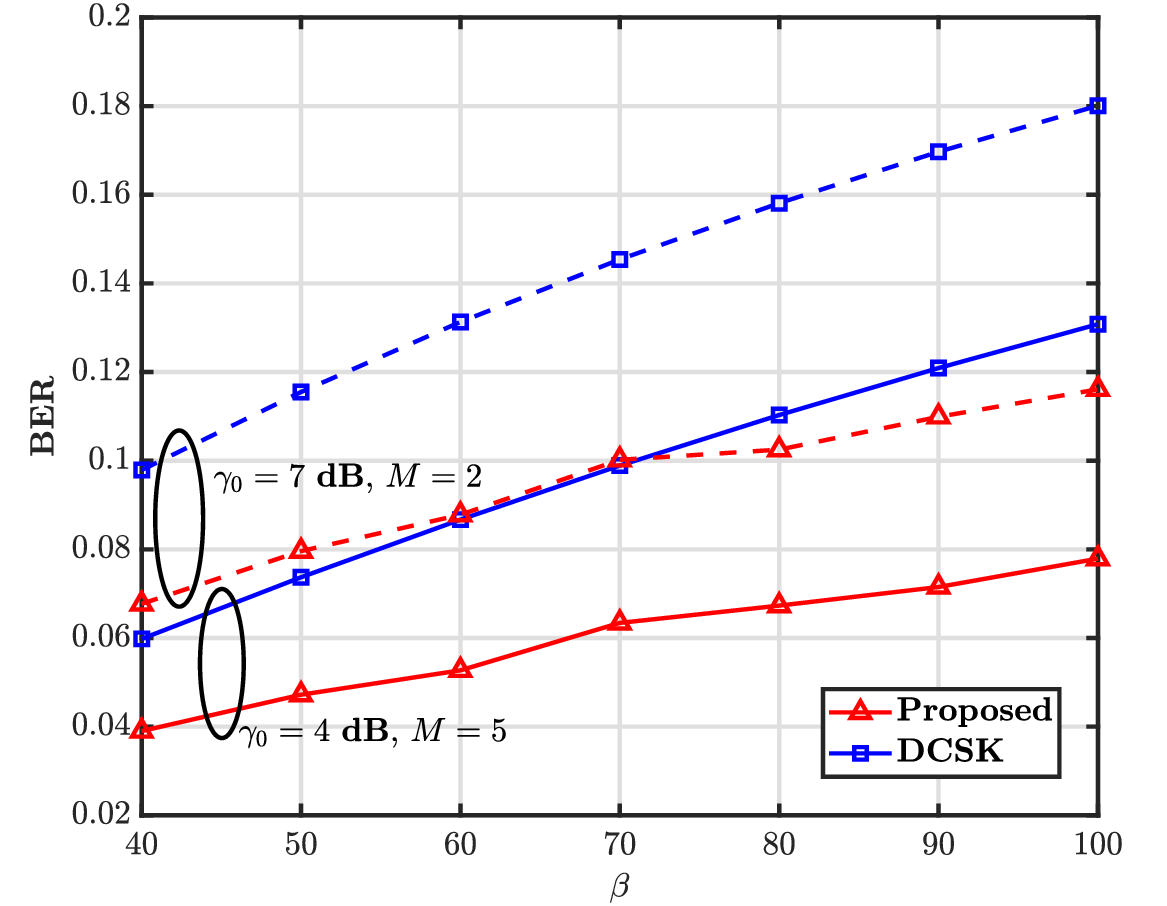}
\vspace{-2mm}
\caption{Impact of proposed and conventional DCSK waveform on BER.}
\label{fig:res6}
\vspace{-2mm}
\end{figure}

Fig. \ref{fig:res6} compares the BER performance of the proposed waveform with that of the conventional DCSK in an AWGN scenario, where we consider the case of $\theta_{e,i}=\theta_{e,j}$ $\forall$ $i,j=1,\dots,M$. Specifically, we look at the two cases of $\gamma_0=7$ dB, $M=2$ and $\gamma_0=4$ dB, $M=5$. In both these cases, we observe that the BER increases with the spreading factor $\beta$ and that the proposed waveform significantly outperforms DCSK. Note that, when $\beta<70$, the BER performance of DCSK corresponding to $\gamma_0=4$ dB, $M=5$ is better compared to the proposed waveform with $\gamma_0=7$ dB, $M=2$. Hence, we can rightly state, that always modulating the DCSK waveform with any arbitrary choice of parameters does not necessarily guarantee an enhanced BER  performance. In the proposed architecture, the decision of modulating the DCSK waveform (or not) should be based on parameters such as $\beta,\gamma_0$, and $M$.

\begin{figure}[!t]
 \centering\includegraphics[width=\linewidth]{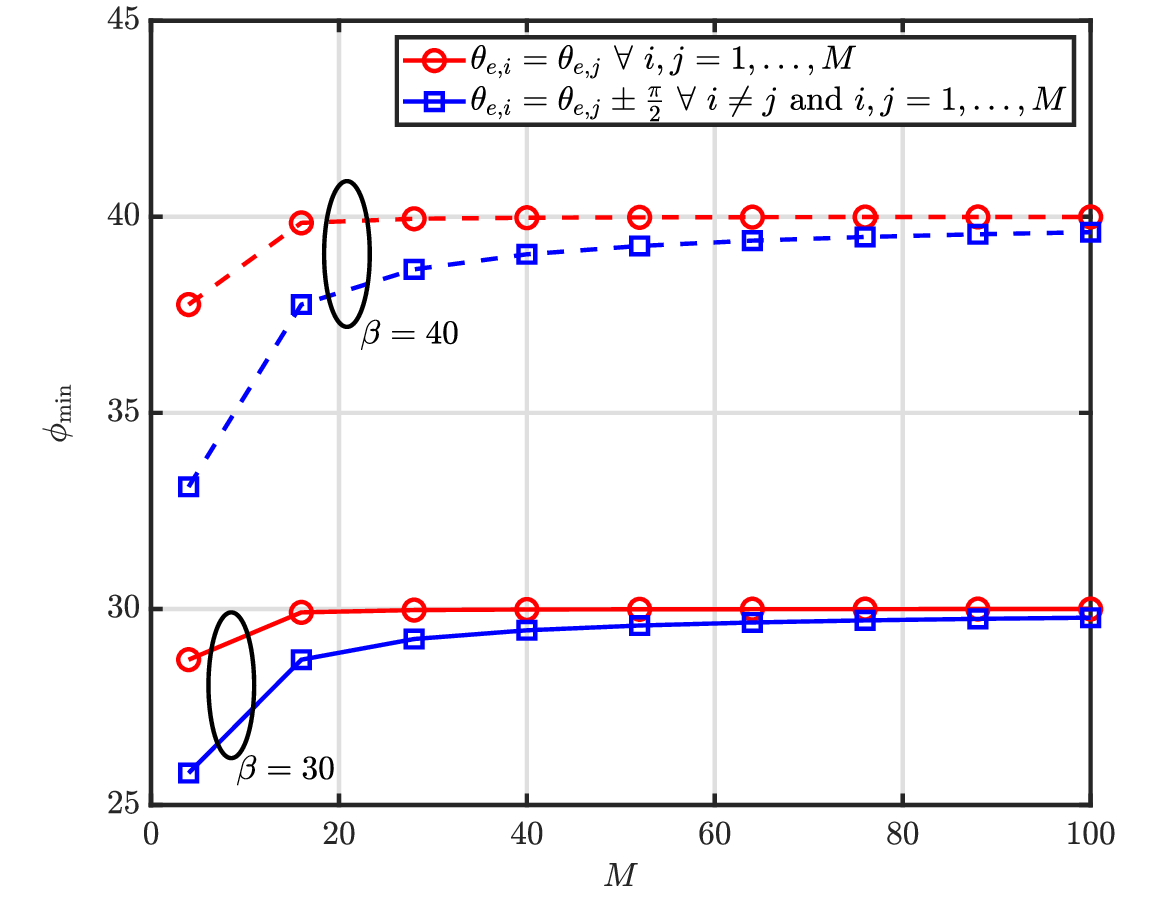}
\vspace{-2mm}
\caption{Impact of $M$ on $\phi_M$; $\gamma_0=16$ dB.}
\label{fig:res3}
\vspace{-2mm}
\end{figure}

Fig. \ref{fig:res3} depicts the impact of $M$ on the optimal reference length $\phi_{\min}$, in an AWGN scenario, for a given $\gamma_0$. The figure illustrates that, in both the cases of $\theta_{e,i}=\theta_{e,j}$ $\forall$ $i,j=1,\dots,M$ and $\theta_{e,i}=\theta_{e,j} \pm \frac{\pi}{2}$ $\forall$ $i \neq j$ and $i,j=1,\dots,M$, $\phi_{\min}$ asymptotically tends to $\beta$ with increasing $M$. This supports the claim made in Corollary \ref{corr1}, where we observed that the advantage of SR-DCSK over DCSK, in terms of reduced frame length, is lost as $\phi_{\min} \rightarrow \beta$ when $M \rightarrow \infty$. In other words, to minimize BER by reducing the frame length, it is not always beneficial to employ more and more reflecting elements for IT.

\begin{figure}[!t]
 \centering\includegraphics[width=\linewidth]{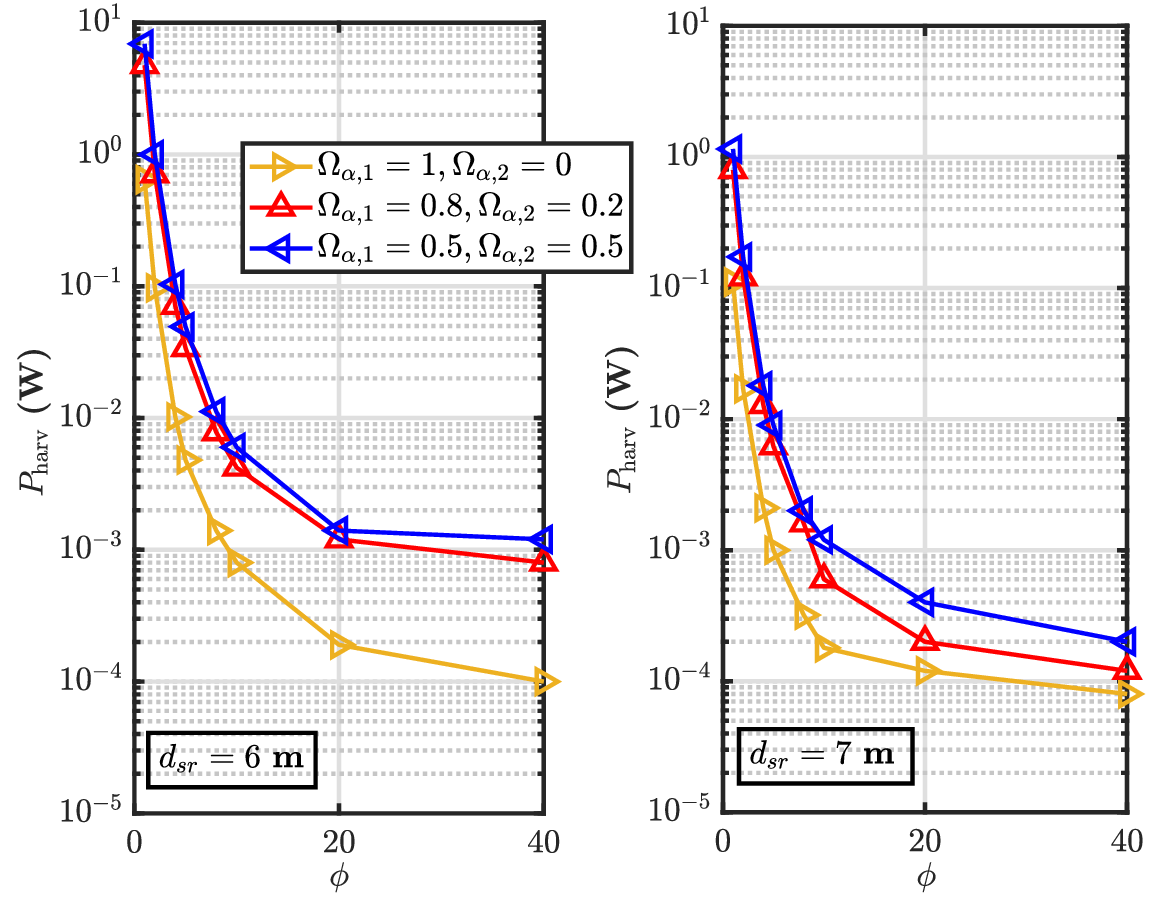}
\vspace{-2mm}
\caption{Impact of $\phi$ on $P_{\rm harv}$; $m=4,\beta=40,$ and $K=20$.}
\label{fig:res4}
\vspace{-2mm}
\end{figure}

In Fig. \ref{fig:res4}, for various degrees of frequency-selectivity, we investigate the impact of the SR-DCSK frame length on the harvested DC power at the RIS, where lines and markers correspond to analytical and simulation results, respectively. We observe that, the case of $\Omega_{\alpha,2}=0$, i.e., the flat fading scenario results in the worst EH performance at the RIS. Moreover, $P_{\rm harv}$ increases as the `degree of frequency-selectivity' of the Tx-RIS channel increases. The figure also demonstrates that having a longer reference leads to a deteriorating WPT performance at the RIS, which corroborates the claim made in \cite{jstsp}. In other words, we observe that the worst $P_{\rm harv}$ is obtained at $\phi=\beta$, i.e., for the case of conventional DCSK. Furthermore, note that even increasing $d_{sr}$ by $1$ m results in significant degradation of $P_{\rm harv}$, which is intuitive because of the additional path-loss factor. Hence, for identical energy requirement at the IT section of the RIS, larger $d_{sr}$ implies higher $K_{\min}$ in \eqref{kmindef}.

\begin{figure}[!t]
 \centering\includegraphics[width=\linewidth]{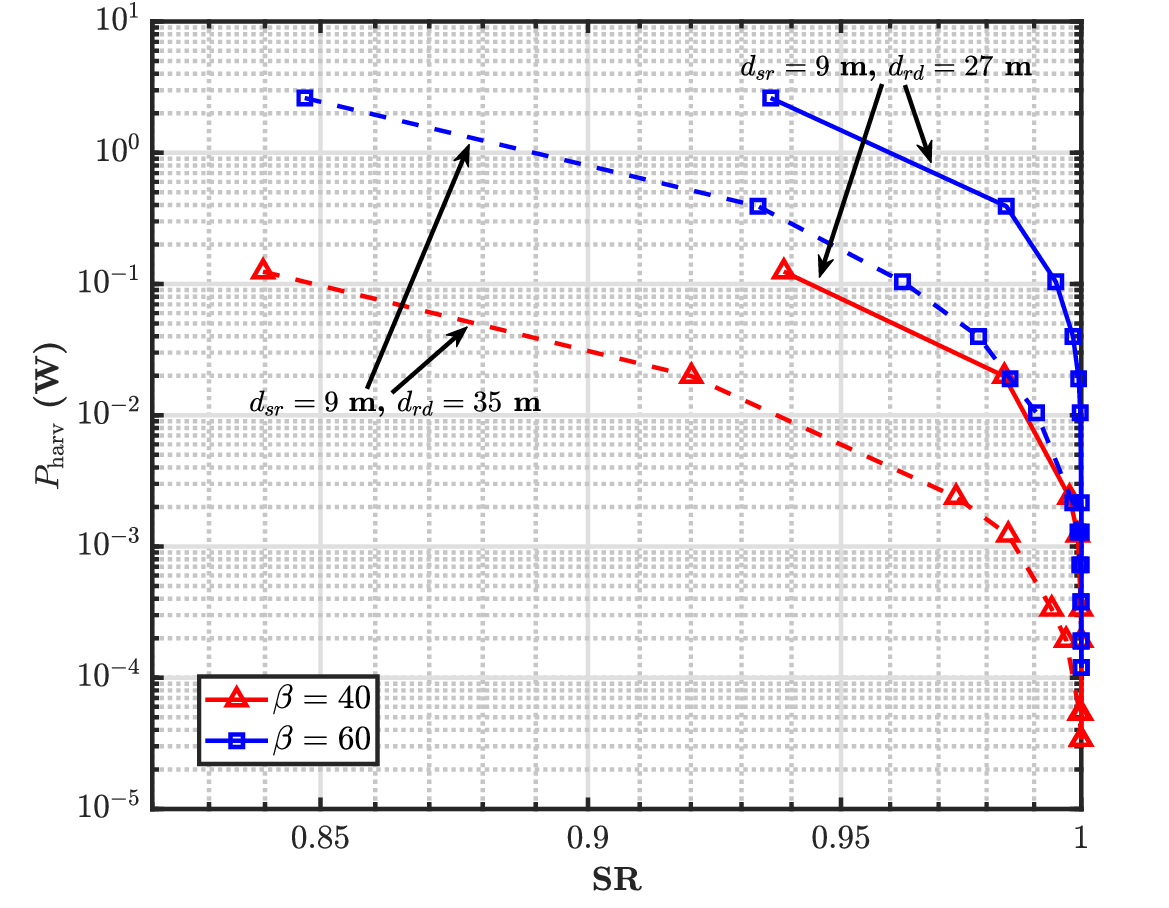}
\vspace{-2mm}
\caption{Impact of system parameters on ${\rm SR}-P_{\rm harv}$ region.}
\label{fig:res5}
\vspace{-2mm}
\end{figure}

Fig. \ref{fig:res5} illustrates the proposed ${\rm SR}-P_{\rm harv}$ region for the considered system with $N=100,M=40,$ and $K=60$. Specifically, we consider spreading factor $\beta=40$ and $\beta=60$ in a Nakagami-$m$ fading scenario with $m=4,\Omega_{\alpha,1}=\Omega_{\beta,1}=0.8,$ and $\Omega_{\alpha,2}=\Omega_{\beta,2}=0.2$ for two cases: $d_{sr}=9$ m, $d_{rd}=27$ m and $d_{sr}=9$ m, $d_{rd}=35$ m. We observe from the figure, that for a fixed $(M,K)$ pair, SR increases with $\phi$ in $[1,\phi_{\min}]$, while the harvested power decreases, i.e., a trade-off. Moreover, under identical scenarios, both SR and $P_{\rm harv}$ corresponding to $\beta=60$ are greater compared to their $\beta=40$ counterpart, which supports the claims made in Theorem \ref{theo1} and Theorem \ref{theo3}. Furthermore, the figure demonstrates that for a given $\beta$, $P_{\rm harv}$ remains identical for both the cases of $d_{sr}=9$ m, $d_{rd}=27$ m and $d_{sr}=9$ m, $d_{rd}=35$ m, while SR deteriorates in the later. The reason for this is attributed to the fact that $P_{\rm harv}$ depends on only $d_{sr}$, while SR depends on both $d_{sr}$ and $d_{rd}$.

\begin{figure}[!t]
 \centering\includegraphics[width=\linewidth]{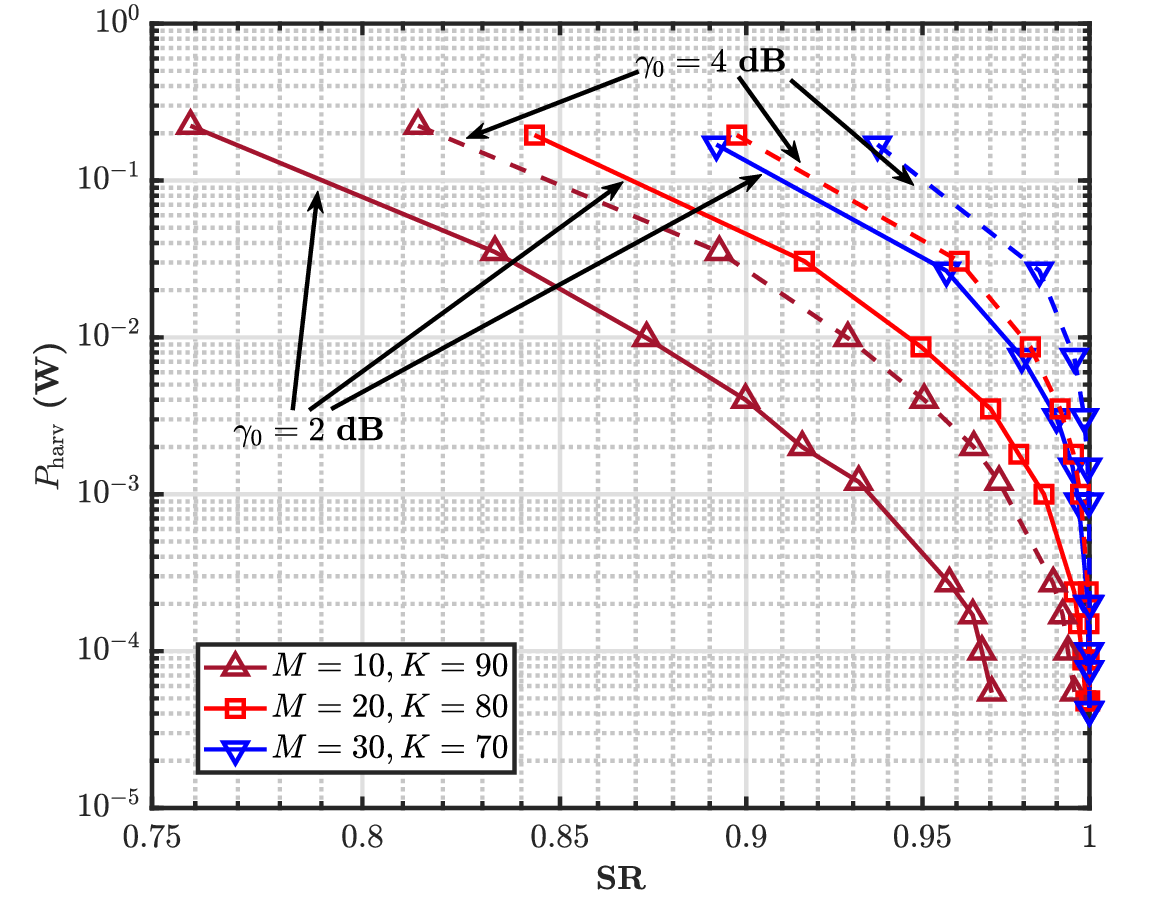}
\vspace{-2mm}
\caption{Impact of RIS configuration on ${\rm SR}-P_{\rm harv}$ region.}
\label{fig:risconfig}
\vspace{-2mm}
\end{figure}

Fig. \ref{fig:risconfig} demonstrates the impact of RIS configuration on the proposed system model, where we consider the wireless channel with $m=8,\Omega_{\alpha,1}=\Omega_{\beta,1}=0.6,\Omega_{\alpha,2}=\Omega_{\beta,2}=0.4$ and $d_{sr}=12$ m, $d_{rd}=40$ m. By taking $N=100$ and $\beta=60$, for both $\gamma_0=2$ dB and $\gamma_0=4$ dB, we evaluate the ${\rm SR}-P_{\rm harv}$ region for three different $(M,K)$ pair. We observe, that for a given $\gamma_0$, an increase in $M(K)$ results in improved ${\rm SR} (P_{\rm harv})$ performance, which corroborates the insights obtained from Theorem \ref{theo1} and Theorem \ref{theo3}, respectively. Also, note that, with all other parameters remaining constant, for a given $(M,K)$ pair, a greater $\gamma_0$ results in an enhanced ${\rm SR}-P_{\rm harv}$ region, and vice-versa.

\begin{figure}[!t]
 \centering\includegraphics[width=\linewidth]{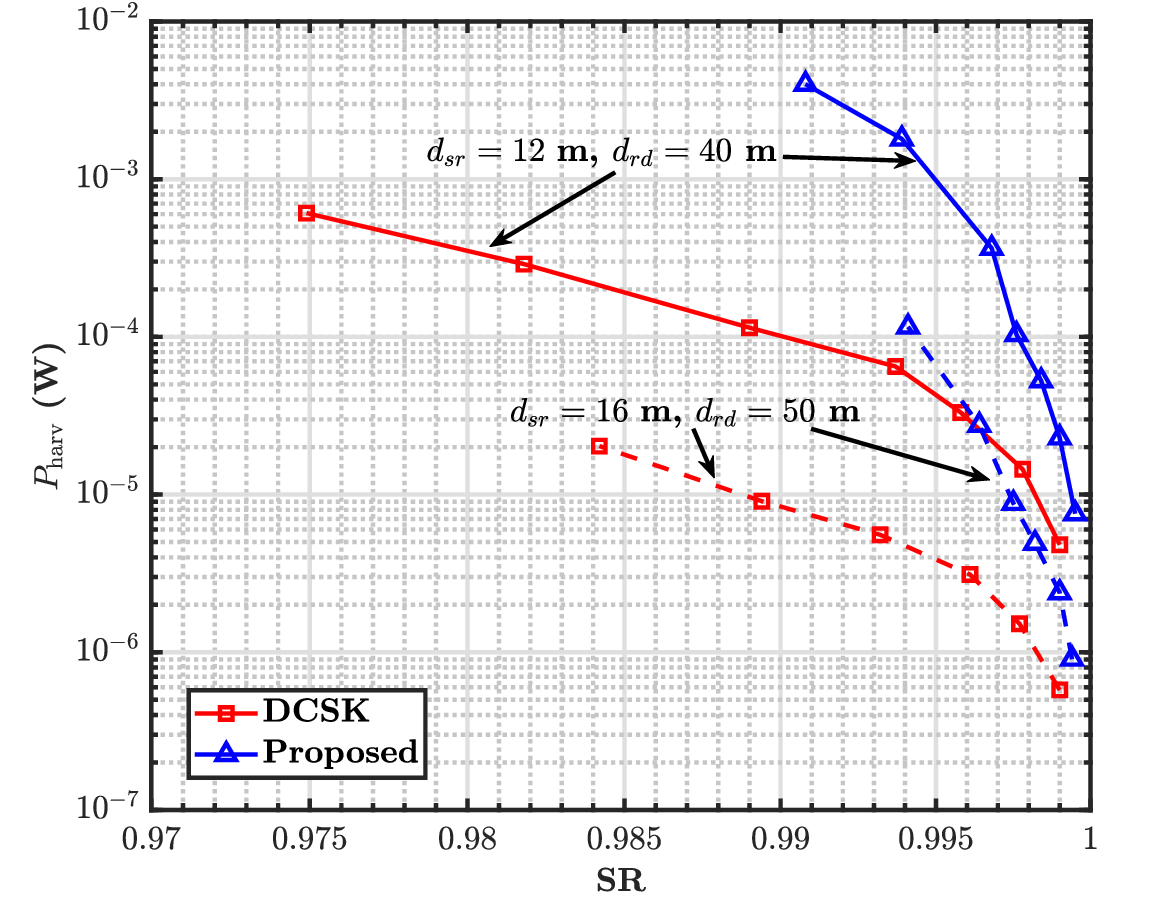}
\vspace{-2mm}
\caption{Comparison of proposed and conventional DCSK waveform.}
\label{fig:regdcsk}
\vspace{-2mm}
\end{figure}

For $N=100$ $(M=20,K=80)$ and $\gamma_0=2$ dB, Fig. \ref{fig:regdcsk} compares the ${\rm SR}-P_{\rm harv}$ region of the proposed and conventional DCSK waveform with the following set of channel parameters: $m=2,\Omega_{\alpha,1}=\Omega_{\beta,1}=0.7,$ and $\Omega_{\alpha,2}=\Omega_{\beta,2}=0.3$. Here we consider $\beta=40,60,80,100,120,160,$ and $200$ to evaluate the reference length $\phi$ for both cases, i.e., we have $\phi=\beta$ for DCSK and evaluate $\phi$ according to Theorem \ref{theo2}. Thereafter, we evaluate the corresponding SR and the harvested energy to obtain the desired ${\rm SR}-P_{\rm harv}$ region. We observe from the figure, that for identical choice of both system parameters and wireless environment, the proposed waveform design results in an enhanced ${\rm SR}-P_{\rm harv}$ region as compared to its conventional DCSK-based counterpart. Moreover, while Fig. \ref{fig:res1}, Fig. \ref{fig:res2}, and Fig. \ref{fig:berch} show the BER performance comparison of the DCSK-based and proposed waveform design, Fig. \ref{fig:res4} demonstrates the WPT performance comparison of these two schemes. Summarizing the claims of all these results together illustrate the enhanced SWIPT performance of the proposed waveform design in comparison to the existing DCSK-based benchmark scheme.  Note that, the justification for this, is the adaptation of the DCSK-based transmit symbol reference length, which in turn, depends on the application specific requirements.

\section{Conclusion}
In this paper, we investigated the effects of conventional communication-based chaotic waveforms in self-sustainable RIS-assisted SWIPT. In particular, we considered a SISO point-to-point set-up, where the transmitter employs a DCSK-based signal generator and the RIS elements are partitioned into two sub-surfaces, namely, the IT and the EH section. While the IT section acts as a relay for the transmitter, the EH section harvests energy from the incident signal, which in turn, is used by the RIS controller and the IT section to attain self-sustainability. By considering a generalized frequency selective wireless channel and taking into account the nonlinearities of the EH process, we characterized the system in terms of the BER and the harvested DC power. We demonstrated that, both these metrics are dependent on the parameters of the transmitter waveform, the wireless channel, as well as the number of reflecting elements in the IT and EH section of the RIS, respectively. Accordingly, we also showed, that it is not always wise to include more and more reflecting elements in the RIS IT section. Moreover, we looked into the BER-harvested energy trade-off and how it impacts the aspect of transmit waveform design.  Furthermore, numerical results illustrate the advantages of the proposed chaotic waveform-based SWIPT architecture. As an immediate extension of this work, we aim to investigate the aspect of considering the RIS IT section as an active device with amplifying factor greater than unity. Moreover, by incorporating the various hardware impairments, other RIS architectures, including simultaneously transmitting and reflecting RIS and beyond diagonal RIS, can also be considered in a DCSK-based multi-user scenario.

\appendices

\section{Proof of Theorem \ref{theo1}}  \label{app1}
By using \eqref{dec}, we obtain
\begin{align}  \label{lamdef}
\lambda_l&= \Re \left(  \sum\limits_{b=1}^{\zeta}\sum\limits_{z=0}^{\phi-1} \Bigg( \delta\sum\limits_{p=1}^{L_{sr}}\sum\limits_{q=1}^{L_{rd}} \sum\limits_{m=1}^M e^{j\theta_{e,m}}\alpha_{p,m}\beta_{m,q}x_{l,z}d_l \right. \nonumber \\
&\left. +w_{b,z+\phi} \Bigg) \!\! \left( \!\delta\sum\limits_{p=1}^{L_{sr}}\sum\limits_{q=1}^{L_{rd}} \sum\limits_{m=1}^M e^{j\theta_{e,m}}\alpha_{p,m}\beta_{m,q}x_{l,z}+w_z \! \right)^*\! \right) \nonumber \\
&\overset{(a)}{=}\Re \left( \! \sum\limits_{b=1}^{\zeta}\sum\limits_{z=0}^{\phi-1} \left( \! \delta^2 \sum\limits_{p=1}^{L_{sr}}\sum\limits_{q=1}^{L_{rd}} \Big|\!\!\sum\limits_{m=1}^M e^{j\theta_{e,m}}\alpha_{p,m}\beta_{m,q}\Big|^2x_{l,z}^2d_l \! \right)\!\! \right) \nonumber\\
& + \Re \left( \sum\limits_{b=1}^{\zeta}\sum\limits_{z=0}^{\phi-1} \left( \delta\sum\limits_{p=1}^{L_{sr}}\sum\limits_{q=1}^{L_{rd}} \sum\limits_{m=1}^M e^{j\theta_{e,m}}\alpha_{p,m}\beta_{m,q}x_{l,z}d_l \! \right)w_z^* \!\right) \nonumber \\
&+ \Re \Bigg( \!\sum\limits_{b=1}^{\zeta}\sum\limits_{z=0}^{\phi-1} w_{b,z+\phi} \nonumber \\
& \times \left( \! \delta\sum\limits_{p=1}^{L_{sr}}\sum\limits_{q=1}^{L_{rd}} \sum\limits_{m=1}^M e^{j\theta_{e,m}}\alpha_{p,m}\beta_{m,q}x_{l,z} \right)^* + w_{b,z+\phi}w_z^* \Bigg) \nonumber \\
&=\zeta\phi\delta^2 \sum\limits_{p=1}^{L_{sr}}\sum\limits_{q=1}^{L_{rd}} \Big|\sum\limits_{m=1}^Me^{j\theta_{e,m}}\alpha_{p,m}\beta_{m,q}\Big|^2x_{l,z}^2d_l \nonumber \\
& + \Re \left( \zeta \sum\limits_{z=0}^{\phi-1} w_z^* \left( \delta\sum\limits_{p=1}^{L_{sr}}\sum\limits_{q=1}^{L_{rd}} \sum\limits_{m=1}^M e^{j\theta_{e,m}}\alpha_{p,m}\beta_{m,q}x_{l,z}d_l \right)\!\!\right) \nonumber \\
&+ \Re \Bigg( \sum\limits_{b=1}^{\zeta}\sum\limits_{z=0}^{\phi-1} w_{b,z+\phi} \nonumber\\
& \times \left(  \delta\sum\limits_{p=1}^{L_{sr}}\sum\limits_{q=1}^{L_{rd}} \sum\limits_{m=1}^M e^{j\theta_{e,m}}\alpha_{p,m}\beta_{m,q}x_{l,z} \right)^* + w_{b,z+\phi}w_z^* \Bigg).
\end{align}
Here, $(a)$ follows from considering the aspect of low cross-correlation of two chaotic sequences, i.e., from \eqref{corprop}, we obtain
\begin{equation}  \label{lcross}
\sum\limits_{z=0}^{\phi-1}x_{l,z_a}x_{l,z_b} \approx 0  \quad {\rm for} \quad a \neq b.
\end{equation}
Moreover, it is to be noted that the noise component $w_{z}$ remains unchanged over all the $\zeta$ partial correlations.

In order to evaluate the BER, in terms of both the system and channel parameters, we need to obtain the mean and variance of the decision variable $\lambda_l$ from \eqref{lamdef}.
\begin{align}  \label{mean}
\mathbb{E}\{ \lambda_l \}&=\mathbb{E} \left\lbrace  \zeta\phi\delta^2 \sum\limits_{p=1}^{L_{sr}}\sum\limits_{q=1}^{L_{rd}} \Big|\sum\limits_{m=1}^Me^{j\theta_{e,m}}\alpha_{p,m}\beta_{m,q}\Big|^2x_{l,z}^2d_l \right.\nonumber \\
& + \left. \Re \left( \!\!\zeta \! \sum\limits_{z=0}^{\phi-1} w_z^* \!\left( \! \delta\sum\limits_{p=1}^{L_{sr}}\sum\limits_{q=1}^{L_{rd}} \sum\limits_{m=1}^M e^{j\theta_{e,m}}\alpha_{p,m}\beta_{m,q}x_{l,z}d_l \right)\!\!\right)\right. \nonumber \\
&+ \left. \Re \left( \sum\limits_{b=1}^{\zeta}\sum\limits_{z=0}^{\phi-1} w_{b,z+\phi} \right.\right. \nonumber \\
& \left.\left.\times \left( \! \delta \! \sum\limits_{p=1}^{L_{sr}}\sum\limits_{q=1}^{L_{rd}} \sum\limits_{m=1}^M \! e^{j\theta_{e,m}}\alpha_{p,m}\beta_{m,q}x_{l,z} \!\!\right)^* \!\!\!+ w_{b,z+\phi}w_z^* \right) \!\!\right\rbrace \nonumber \\
&\overset{(a)}{=} \zeta\phi P_{\rm t}C_0^2d_{sr}^{-\alpha_{sr}}d_{rd}^{-\alpha_{rd}} \mathbb{E} \{ x_k^2 \} d_l \nonumber \\
& \qquad\qquad\times \sum\limits_{p=1}^{L_{sr}}\sum\limits_{q=1}^{L_{rd}} \Big|\sum\limits_{m=1}^Me^{j\theta_{e,m}}\alpha_{p,m}\beta_{m,q}\Big|^2 \nonumber \\
&\overset{(b)}{=}\frac{\beta}{\left( \beta+\phi \right)}\gamma_0N_0 d_l \sum\limits_{p=1}^{L_{sr}}\sum\limits_{q=1}^{L_{rd}} \Big|\sum\limits_{m=1}^Me^{j\theta_{e,m}}\alpha_{p,m}\beta_{m,q}\Big|^2,
\end{align}
where $(a)$ follows from the fact that all except the first term have zero mean. Moreover, $(b)$ follows from $\beta=\zeta\phi$ and by using \eqref{rsig1}, \eqref{biten}, and \eqref{gamdef}, respectively. Hereafter, based on the independence property of random variables, we obtain
\begin{align}  \label{var}
{\rm Var}\{\lambda_l\}&={\rm Var} \left\lbrace \Re \left( \zeta \sum\limits_{z=0}^{\phi-1} w_z^* \right.\right. \nonumber \\
& \!\!\!\!\!\!\!\!\!\!\!\!\!\!\!\!\left.\left.\times \left( \delta\sum\limits_{p=1}^{L_{sr}}\sum\limits_{q=1}^{L_{rd}} \sum\limits_{m=1}^M e^{j\theta_{e,m}}\alpha_{p,m}\beta_{m,q}x_{l,z}d_l \right)\right) \right. \nonumber \\
& \!\!\!\!\!\!\!\!\!\!\!\!\!\!\!\! + \left.\Re \left( \sum\limits_{b=1}^{\zeta}\sum\limits_{z=0}^{\phi-1} w_{b,z+\phi} \right.\right. \nonumber \\
& \!\!\!\!\!\!\!\!\!\!\!\!\!\!\!\!\left. \left. \times \left(  \delta\sum\limits_{p=1}^{L_{sr}}\sum\limits_{q=1}^{L_{rd}} \sum\limits_{m=1}^M e^{j\theta_{e,m}}\alpha_{p,m}\beta_{m,q}x_{l,z} \right)^* + w_{b,z+\phi}w_z^* \right)\right\rbrace \nonumber \\
&\!\!\!\!\!\!\!\!\!\!\!\!\!\!\!\!=\zeta^2\phi\delta^2 \frac{N_0}{2} \mathbb{E} \{ x_k^2 \} \sum\limits_{p=1}^{L_{sr}}\sum\limits_{q=1}^{L_{rd}} \Big|\sum\limits_{m=1}^M e^{j\theta_{e,m}}\alpha_{p,m}\beta_{m,q}\Big|^2 \nonumber \\
&\!\!\!\!\!\!\!\!\!\!\!\!\!\!\!\!+ \zeta\phi\delta^2 \frac{N_0}{2} \mathbb{E} \{ x_k^2 \} \sum\limits_{p=1}^{L_{sr}}\sum\limits_{q=1}^{L_{rd}} \Big|\sum\limits_{m=1}^M e^{j\theta_{e,m}}\alpha_{p,m}\beta_{m,q}\Big|^2 \!\! + \zeta\phi\frac{N_0^2}{4} \nonumber \\
&\!\!\!\!\!\!\!\!\!\!\!\!\!\!\!\!\overset{(a)}{=}\frac{\beta N_0^2}{2} \left( \sum\limits_{p=1}^{L_{sr}}\sum\limits_{q=1}^{L_{rd}} \Big|\!\sum\limits_{m=1}^M e^{j\theta_{e,m}}\alpha_{p,m}\beta_{m,q}\Big|^2 \frac{\gamma_0 \left( \zeta+1 \right)}{\beta+\phi} \! + \! \frac{1}{2} \!\! \right),
\end{align}
where $(a)$ follows from using $\beta=\zeta\phi$, \eqref{rsig1}, \eqref{biten}, and \eqref{gamdef}, respectively. Note that, as both quantities $\lambda_l |(d_l=-1)$ and $\lambda_l |(d_l=+1)$ are sum of a large number of random variables, we assume them to be Gaussian distributed \cite{text}. Hence, by assuming equally probable transmission of $d_l=\pm 1$, the system BER (based on channel conditions, i.e., $\alpha_{p,m},\beta_{m,q}$ $\forall$ $p=1,\dots,L_{sr},q=1,\dots,L_{rd},$ and $m=1,\dots,M$) is
\begin{align} \label{ber}
{\rm BER}&=\frac{1}{2}\mathbb{P} \left\lbrace \lambda_l<1 | d_l=+1 \right\rbrace  + \frac{1}{2}\mathbb{P} \left\lbrace \lambda_l>1 | d_l=-1 \right\rbrace \nonumber \\
&=\frac{1}{2}{\rm erfc} \left( \left[\frac{2{\rm Var}\{\lambda_l| d_l=+1\}}{\mathbb{E}^2\{\lambda_l| d_l=+1\}} \right]^{-\frac{1}{2}}   \right).
\end{align}
Hence, by using \eqref{mean} and \eqref{var} in this definition, we obtain the conditional BER as
\begin{align}
&{\rm BER}= \nonumber \\
&\frac{1}{2}{\rm erfc} \! \left( \!\! \left[\!\frac{ \sum\limits_{p=1}^{L_{sr}}\sum\limits_{q=1}^{L_{rd}} \Big|\sum\limits_{m=1}^M e^{j\theta_{e,m}}\alpha_{p,m}\beta_{m,q}\Big|^2 \frac{\gamma_0 \left( \zeta+1 \right)}{\beta+\phi} + \frac{1}{2}}{\frac{\beta\gamma_0^2}{\left( \beta+\phi \right)^2} \left(\sum\limits_{p=1}^{L_{sr}}\sum\limits_{q=1}^{L_{rd}} \Big|\sum\limits_{m=1}^Me^{j\theta_{e,m}}\alpha_{p,m}\beta_{m,q}\Big|^2\right)^2}\! \right]^{-\frac{1}{2}} \!  \right).
\end{align}
Accordingly, we define $\Lambda=\sum\limits_{p=1}^{L_{sr}}\sum\limits_{q=1}^{L_{rd}} \Big|\sum\limits_{m=1}^M e^{j\theta_{e,m}}\alpha_{p,m}\beta_{m,q}\Big|^2$, which after trivial algebraic manipulations result in
\begin{equation}  \label{berdef}
{\rm BER}(\Lambda)=\frac{1}{2} {\rm erfc} \left( \left[ \frac{\left( 1+\zeta \right)^2}{\gamma_0\zeta\Lambda} + \frac{\left( \beta+\phi \right)^2}{2\beta\gamma_0^2\Lambda^2}\right]^{-\frac{1}{2}}   \right).
\end{equation}
Therefore, we obtain
\begin{equation}
{\rm BER}=\int\limits_0^{\infty} {\rm BER}(\Lambda) f(\Lambda) d\Lambda,
\end{equation}
where $f(\Lambda)$ denotes the probability density function of $\Lambda$. Note that, the probability density function for $\Big|\sum\limits_{m=1}^M e^{j\theta_{e,m}}\alpha_{p,m}\beta_{m,q}\Big|$ is investigated in \cite{dnaka} for a Nakagami-$m$ fading scenario. Hence, we can evaluate $f(\Lambda)$ by the standard technique of transformation of random variables \cite{papoulis}.

\section{Proof of Theorem \ref{theo2}}  \label{app2}
Due to the AWGN assumption, in \eqref{bertheo}, we replace $\alpha_{1,m}=\beta_{m,1}=1$ and $\alpha_{p,m}=\beta_{m,q}=0$ $\forall$ $p \neq 1$, $q \neq 1$, and $m=1,\dots,M$  to obtain
\begin{equation}  \label{berawgn}
{\rm BER}=\frac{1}{2} {\rm erfc} \left( \left[ \frac{\left( 1+\zeta \right)^2}{\gamma_0\zeta\Lambda} + \frac{\left( \beta+\phi \right)^2}{2\beta\gamma_0^2\Lambda^2}\right]^{-\frac{1}{2}}   \right).
\end{equation}
Moreover, the quantity $\Lambda$ defined in \eqref{berawgn} simplifies as follows.
\begin{align}
\Lambda&=\Big|\sum\limits_{m=1}^M e^{j\theta_{e,m}}\Big|^2 \overset{(a)}{=} \left( \sum\limits_{m=1}^M \cos \theta_{e,m} \right)^2 + \left( \sum\limits_{m=1}^M \sin \theta_{e,m} \right)^2 \nonumber \\
& \overset{(b)}{=} M+ 2\sum_{\substack{i,j=1 \\ i \neq j}}^M \cos \left( \theta_{e,i}-\theta_{e,j} \right),
\end{align}
where $(a)$ follows from using $e^{j\theta}=\cos \theta + j \sin \theta$ and $(b)$ follows from using the trigonometric identity $\sin^2 \theta+ \cos^2 \theta=1$ and $\cos (x) =\cos (-x)$ $\forall$ $x$. Hence, we obtain
\begin{align}
{\rm BER}&=\frac{1}{2} {\rm erfc} \left(  \left[ \frac{\left( 1+\zeta \right)^2}{\gamma_0\zeta \Lambda} + \frac{\left( \beta+\phi \right)^2}{2\beta\gamma_0^2\Lambda^2}\right]^{-\frac{1}{2}}   \right) \nonumber \\
&=\frac{1}{2} {\rm erfc} \left(  \frac{1}{\sqrt{\Psi(\phi)}}   \right),
\end{align}
where we define
\begin{equation}
\Psi(\phi)=\frac{\left( 1+\zeta \right)^2}{\gamma_0\zeta \Lambda} + \frac{\left( \beta+\phi \right)^2}{2\beta\gamma_0^2\Lambda^2}.
\end{equation}
As our objective is to obtain the reference length with minimum BER, we look into the first derivative of $\Psi(\phi)$ with respect to $\phi$, i.e., $\Psi'(\phi)=\frac{\partial\Psi(\phi)}{\partial\phi}$. By using $\beta=\zeta\phi$, we rewrite $\Psi(\phi)$ as
\begin{equation}
\Psi(\phi)= \frac{\left( 1+\zeta \right)^2}{\gamma_0\zeta \Lambda} + \frac{\left( \beta+\phi \right)^2}{2\beta\gamma_0^2\Lambda^2}=\frac{\left( \beta+\phi \right)^2}{\gamma_0\Lambda\beta} \left( \frac{1}{\phi} + \frac{1}{2\gamma_0\Lambda} \right).
\end{equation}
Accordingly, by putting $\Psi'(\phi)=0$ and taking into account $\phi>0$ and $\gamma_0 \in \mathbb{R}^+$, we obtain the corresponding critical point as
\begin{equation}
\phi_{\rm critical}=\frac{\gamma_0\Lambda}{2} \left( \sqrt{1+\frac{4\beta}{\gamma_0\Lambda}}-1 \right).
\end{equation}
In order to decide on $\phi_{\rm critical}$ being a maxima or minima, we evaluate $\Psi''(\phi)=\frac{\partial^2\Psi(\phi)}{\partial\phi^2}$ at $\phi_{\rm critical}$. It turns out that
\begin{equation}
\Psi''(\phi_{\rm critical})=\frac{1}{\beta \gamma_0^2 \Lambda^2} + \frac{2\beta}{\phi_{\rm critical}^3\gamma_0\Lambda} >0.
\end{equation}
Hence, the minimum value of $\Psi(\phi)$ is obtained at $\phi_{\min}=\phi_{\rm critical}$. Therefore, based on the nature of ${\rm erfc}(x)$ for $x \geq 0$, we conclude that for a given set of system parameters, the minimum BER is attained for $\phi=\phi_{\min}$.

\section{Proof of Theorem \ref{theo3}}  \label{app3}
By using the channel model considered in Section \ref{chmodel} and the assumption of identical channel statistics across the RIS elements, we now evaluate $\mathbb{E}\{ \mathcal{H}_i^2 \}$ and $\mathbb{E}\{ \mathcal{H}_i^4 \}$, respectively.
\begin{align}  \label{h2}
\mathbb{E}\left\lbrace \mathcal{H}_i^2 \right\rbrace &=\mathbb{E}\left\lbrace  \sum\limits_{p=1}^{L_{sr}} \alpha_{p,i}^2 + 2 \sum_{\mathclap{\substack{p_1,p_2=1\\p_1 \neq p_2}}}^{L_{sr}} \alpha_{p_1,i}\alpha_{p_2,i}\right\rbrace \nonumber \\
& \!\!\!\!\!\!\!\! =1+\frac{2}{m_s}\left(\frac{\Gamma(m_s+0.5)}{\Gamma(m_s)}\right)^2 \sum_{\mathclap{\substack{p_1,p_2=1\\p_1 \neq p_2}}}^{L_{sr}} \sqrt{\Omega_{\alpha,p_1}\Omega_{\alpha,p_2}}.
\end{align}
Here we use the $n$-th order moment of $\alpha_{p,i}$, i.e.,
\begin{align}    \label{val1}
\mathbb{E}\{\alpha_{p,i}^n\}&=\frac{2m_s^{m_s}}{\Gamma(m_s)\Omega_{\alpha,p}^{m_s}}\int_0^{\infty} z^{2m_s+n-1}e^{-\frac{m_sz^2}{\Omega_{\alpha,p}}} dz \nonumber \\
& =\frac{\Gamma(m_s+\frac{n}{2})}{\Gamma(m_s)}\left(\frac{\Omega_{\alpha,p}}{m_s}\right)^{\frac{n}{2}}.
\end{align}
This follows from the transformation $ \frac{m_sz^2}{\Omega_{\alpha,p}} \rightarrow \omega$.
Based on \eqref{val1} and by using the multinomial theorem, we obtain
\begin{align}  \label{h4}
&\mathbb{E}\left\lbrace \mathcal{H}_i^4 \right\rbrace= \mathbb{E}\left\lbrace \sum\limits_{k_1+k_2+\dots+k_{L_{sr}}=4}\! \frac{4!}{k_1! \: k_2! \: \dots \: k_{L_{sr}}!}\! \prod\limits_{p=1}^{L_{sr}} \alpha_{p,i}^{k_p} \right\rbrace\nonumber \\
&=\!\!\!\!\sum\limits_{k_1+k_2+\dots+k_{L_{sr}}=4}\! \frac{4!}{k_1! \: k_2! \: \dots \: k_{L_{sr}}!}\! \prod\limits_{p=1}^{L_{sr}} \! \frac{\Gamma(m_s+\frac{k_p}{2})}{\Gamma(m_s)}\!\!\left(\!\frac{\Omega_{\alpha,p}}{m_s}\!\right)^{\!\frac{k_p}{2}}\!\!\!\!.
\end{align}
Next, based on the SR-DCSK frame construction in \eqref{symsr}, we have $\mathcal{S}=\sum\limits_{k=1}^{\phi+\beta} s_{l,k}=\left( 1+\zeta d_l \right)\sum\limits_{k=1}^{\phi}x_k$, where the invariant probability density function (PDF) of $x_k$, is \cite{text}
\begin{align}  \label{spdf}
f_X(x)=\begin{cases} 
\dfrac{1}{\pi\sqrt{1-x^2}}, & |x|< 1,\\
0, & \text{otherwise}.
\end{cases}&
\end{align}
Accordingly, by assuming equally likely transmission of information bits $d_l=\pm 1$ and by using \eqref{lcross} and \eqref{spdf}, we obtain
\begin{align}  \label{s2}
\mathbb{E}\left\lbrace \mathcal{S} ^2 \right\rbrace & = \mathbb{E}\left\lbrace \left( 1+\zeta d_l \right)^2 \left( \sum\limits_{k=1}^{\phi}x_k \right) ^2 \right\rbrace \nonumber \\
& = \left(  1+\zeta^2 \right) \!\! \sum\limits_{k=1}^{\phi}\int_{-1}^{1}\frac{x_k^2dx_k}{\pi\sqrt{1-x_k^2}}=\frac{\left(  1+\zeta^2 \right)}{2}\phi.
\end{align}
Moreover, by using \eqref{lcross} and the multinomial framework as demonstrated above, we obtain
\begin{align}  \label{s4}
\mathbb{E}\left\lbrace \mathcal{S} ^4 \right\rbrace&= \mathbb{E}\left\lbrace \left( 1+\zeta d_l \right)^4 \left( \sum\limits_{k=1}^{\phi}x_k \right) ^4 \right\rbrace \nonumber \\
& =\left( 1+6\zeta^2+\zeta^4 \right)\!\mathbb{E}\!\left\lbrace \left( \sum\limits_{k=1}^{\phi}x_k \right) ^4 \right\rbrace \nonumber \\
&= \left( 1+6\zeta^2 \!+\!\zeta^4 \right)\!\! \left( \sum\limits_{k=1}^{\phi}\mathbb{E}\left\lbrace x_k^4 \right\rbrace+ \! 3\mathbb{E}^2\left\lbrace x_k^2 \right\rbrace\phi(\phi-1) \!\!\right) \nonumber \\
&=\left( 1+6\zeta^2+\zeta^4 \right) \left( \sum\limits_{k=1}^{\phi} \int_{-1}^{1}\frac{x_k^4dx_k}{\pi\sqrt{1-x_k^2}} \right. \nonumber \\
& \left.+ 3 \left( \int_{-1}^{1}\frac{x_k^2dx_k}{\pi\sqrt{1-x_k^2}} \right)^2\phi(\phi-1) \right) \nonumber \\
&= \left( 1+6\zeta^2+\zeta^4 \right) \left(  \frac{3\phi}{8}+\frac{3\phi(\phi-1)}{4} \right) \nonumber \\
&=\frac{3\phi}{8}\left( 1+6\zeta^2+\zeta^4 \right) \left( 2\phi-1 \right).
\end{align}
Finally, by replacing \eqref{h2}, \eqref{h4}, \eqref{s2}, and \eqref{s4} in  \eqref{pharv}, we obtain \eqref{pharvfinal}.

\bibliographystyle{IEEEtran}
\bibliography{refs}

\end{document}